\documentclass[aps,pra,reprint,superscriptaddress,nofootinbib,floatfix]{revtex4-2}

\usepackage[T1]{fontenc}
\usepackage[utf8]{inputenc}
\usepackage{amsmath}
\usepackage{amssymb}
\usepackage{bm}
\usepackage{booktabs}
\usepackage{graphicx}
\usepackage{microtype}
\usepackage{hyperref}
\usepackage{lmodern}
\usepackage{xcolor}
\usepackage{float}

\hypersetup{
  colorlinks=true,
  linkcolor=blue,
  citecolor=blue,
  urlcolor=blue
}

\newcommand{\ii}{\mathrm{i}}
\newcommand{\ee}{\mathrm{e}}
\newcommand{\one}{\mathbf{1}}

\newcommand{\dd}{\mathrm{d}}
\newcommand{\Bcal}{\mathcal B}

\newcommand{\ket}[1]{\left|#1\right\rangle}
\newcommand{\bra}[1]{\left\langle#1\right|}
\newcommand{\braket}[2]{\left\langle#1\middle|#2\right\rangle}
\newcommand{\pd}{\partial}

\begin{document}

\title{Imaginary magnetic fields for ultracold atoms}

\author{Saulius Alminas}
\email{saulius.alminas@ff.vu.lt}
\affiliation{Institute of Theoretical Physics and Astronomy, Faculty of Physics, Vilnius University, Saul\.{e}tekio 3, LT-10257 Vilnius, Lithuania.}

\author{Domantas Burba}
\email{domantas.burba@ff.vu.lt}
\affiliation{Institute of Theoretical Physics and Astronomy, Faculty of Physics, Vilnius University, Saul\.{e}tekio 3, LT-10257 Vilnius, Lithuania.}

\author{Gediminas Juzeli\={u}nas}
\email{gediminas.juzeliunas@tfai.vu.lt}
\affiliation{Institute of Theoretical Physics and Astronomy, Faculty of Physics, Vilnius University, Saul\.{e}tekio 3, LT-10257 Vilnius, Lithuania.}

\begin{abstract}
We propose a four-level scheme for generating a uniform imaginary magnetic field for
ultracold atoms using Raman coupling and state-dependent particle losses.
Eliminating two lossy auxiliary states and subsequently projecting onto one of the two remaining Raman-dressed atomic states, we find that the atomic motion is affected by an imaginary vector potential which is engineered to have a constant curl corresponding to a spatially uniform imaginary synthetic magnetic field. We analytically derive the Gaussian wavepacket dynamics for atoms in such a synthetic magnetic field and
study the width-dependent transport, unlike motion in a uniform real magnetic field. We find numerical agreement between the two-level non-Hermitian Hamiltonian and the adiabatic projection description, confirming the applicability of the latter adiabatic approach.
\end{abstract}

\maketitle
\raggedbottom

\section{Introduction}
\label{sec:introduction}

Non-Hermitian Hamiltonians provide an effective description of quantum
systems subject to loss and exhibit spectral and transport properties
absent from their Hermitian counterparts~\cite{AshidaGongUeda2020,
BergholtzBudichKunst2021}. Their eigenstates need not be orthogonal, and
both eigenvalues and eigenvectors can coalesce at exceptional
points~\cite{MiriAlu2019}. Spatially nonreciprocal couplings can also
produce boundary accumulation through the non-Hermitian skin
effect~\cite{HatanoNelson1996,YaoWang2018,OkumaSato2020}. These phenomena
show that dissipation can modify quantum motion rather than simply
reduce the number of particles.

Ultracold atoms provide a convenient setting for studying such effects
in the presence of synthetic gauge fields. In Hermitian systems, atoms adiabatically following
a position-dependent dressed state experience (real) geometric vector and
scalar potentials~\cite{DalibardJuzeliunas2011,Goldman2014}. The
corresponding (real)  synthetic magnetic field allows neutral atoms to simulate
the motion of charged particles in a magnetic
field~\cite{Juz05PRA,Juz05JPB}. Raman dressing has been used to realize
synthetic vector potentials, magnetic fields, and spin-orbit
coupling~\cite{LinVector2009,LinMagnetic2009,LinSOC2011}. Spatially
dependent dark states also provide a means of creating subwavelength
geometric potentials and topological optical
lattices~\cite{Zoller2016,Jendrzejewski2016,Gvozdiovas23PRA,Burba25PRR,Nascimbene25PRL}.

Adding state-dependent loss extends these possibilities to
non-Hermitian atomic systems. The combination of coherent  atom-light coupling and
controlled removal  of atoms has been used to observe chiral state
transfer~\cite{RenJo2022} and dynamical signatures of the skin
effect~\cite{LiangYan2022,ZhaoJo2025}. Recently, Tao \emph{et al.}
realized a spatially uniform imaginary  vector potential in a
 Raman-coupled Bose-Einstein
condensate~\cite{TaoSpielman2026}. Such a vector potential has zero curl.
An imaginary magnetic field instead requires a spatially varying
imaginary vector potential with a nonzero curl \cite{MontagOzawa2026}.

Imaginary magnetic flux has been studied in two-dimensional lattice
models, where it changes the wavefunction norm~\cite{OzawaHayata2024}.
Continuum complex magnetic fields and their Landau-level structure have
also been considered~\cite{MontagOzawa2026}. Non-Hermitian adiabatic
evolution can involve complex geometric phases~\cite{GarrisonWright1988}
and geometric vector and scalar potentials~\cite{MontagOzawaGeometry2026}.
These developments motivate  the atom-light coupling scheme  studied here, in which a spatially dependent
imaginary synthetic vector potential and the corresponding imaginary synthetic magnetic field are obtained by adding controlled particle removal.

For  atoms ``dressed'' with laser radiation, the  synthetic magnetic field depends on both the loss profile
and the coherent  atom-light coupling. In the two-level Raman configuration considered
here, the vector potential is governed by  the ratio
$\Delta/\sqrt{\Omega_R^2-\Delta^2}$, where $\Delta$ is the real amplitude
of the imaginary differential detuning. A linear detuning at fixed Raman
amplitude therefore gives a nonuniform field, with a gap closing at
$|\Delta|=\Omega_R$. Moreover,  for spatially dependent gauge potentials, removal  of atoms from only  one of the two Raman-coupled internal states, like in the experiment  \cite{TaoSpielman2026},
introduces a scalar loss gradient that affects the motion independently
of the magnetic field.

Here, we propose a four-level dissipative  scheme that separates
these effects.  Two additional levels provide tuneable losses to each of the two  Raman-coupled states,  leading to a
spatially varying differential  imaginary detuning (differential attenuation).  

A suitable amplitude  of Raman coupling keeps the dressed-state gap constant and yields
a uniform imaginary magnetic field. We derive the projected Hamiltonian
with unit right-state norm, so that its scalar wavefunction directly
determines the physical density. For the resulting quadratic approximation,
we obtain the Gaussian first moments and covariance, identify a transverse
drift governed by differential attenuation, and compare the prediction with
two-level numerical evolution.

The paper is organized as follows. Section~\ref{sec:model} introduces
the coupling scheme and its adiabatic reduction. Section~\ref{sec:gauss} establishes the Gaussian dynamics
and analyzes the center-of-mass motion and squeezing dynamics for a launch along $y$. Section~\ref{sec:sig} considers
imaginary Aharonov-Bohm transport, and Sec.~\ref{sec:validity} tests
the adiabatic approximation. Section~\ref{sec:conclusions} summarizes
the results. The appendices give the general Gaussian solution,
geometric potentials, and analytical derivations.

\section{Model}
\label{sec:model}

\subsection{Four-level coupling scheme}
\label{sec:four-level}

\begin{figure}[htb]
 \centering
  \includegraphics[width=\columnwidth]
   {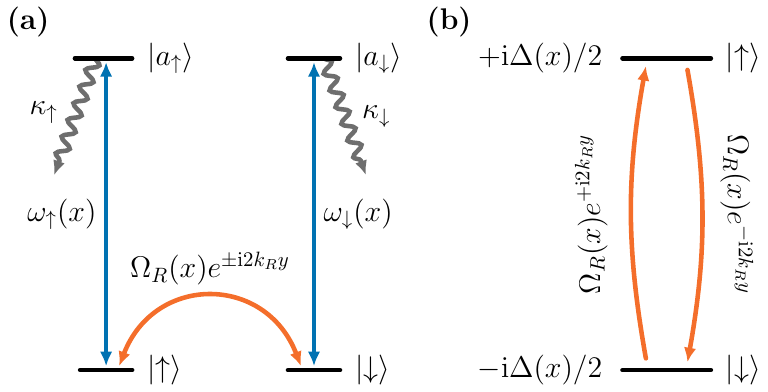}
 \caption{Proposed coupling scheme. (a) Raman-coupled system states
 with independent auxiliary removal channels. The coherent couplings
 are $\Omega_R$ and $\omega_\lambda$, and $\kappa_\lambda$ are
 auxiliary-state decay rates. (b) Effective two-level description
 with purely imaginary detunings $\pm\ii\Delta(x)/2$. The state-independent
 loss term $-\ii\hbar\Gamma (x)/2$ featured in Eq.~\eqref{eq:H2-eff} is not shown here.
}
 \label{fig:loss-scheme}
\end{figure}

We consider two Raman-coupled  atomic internal states $\lvert\uparrow\rangle$ and
$\lvert\downarrow\rangle$, together with two auxiliary states
$\lvert a_\uparrow\rangle$ and $\lvert a_\downarrow\rangle$, as shown in
Fig.~\ref{fig:loss-scheme}. Each  Raman-coupled state $\lvert\lambda\rangle$, with
$\lambda\in\left\{\uparrow,\downarrow\right\}$, is coupled to  the corresponding auxiliary state by radiation characterized by the Rabi frequency $\omega_\lambda$. The  auxiliary states  irreversibly decay
with loss rates $\kappa_\lambda$  through transitions to independent levels outside the present four-level system.
We will not consider these independent levels since there is no population return to the four-level scheme of interest. Thus, the surviving ensemble is described by non-Hermitian
evolution~\cite{DalibardCastinMolmer1992,Daley2014}.

In the four-level basis
$\{\lvert\uparrow\rangle,\lvert\downarrow\rangle,
\lvert a_\uparrow\rangle,\lvert a_\downarrow\rangle\}$, the Hamiltonian
in the rotating frame is
\begin{equation}
\begin{aligned}
 H_4={}&
 H_0\one+\frac{\hbar\Omega_R(x)}{2}\ee^{\ii2k_Ry}\lvert\uparrow\rangle\langle\downarrow\rvert+\mathrm{H.c.}
 \\
 &-\frac{\ii\hbar}{2} \sum_{\lambda\in\left\{\uparrow,\downarrow\right\}}
 \kappa_\lambda
 \lvert a_\lambda\rangle\langle a_\lambda\rvert
 \\
 &+\frac{\hbar}{2}
 \sum_{\lambda\in\left\{\uparrow,\downarrow\right\}} 
 \omega_\lambda(x)\lvert a_\lambda\rangle\langle\lambda\rvert
 +\mathrm{H.c.}\,,
\end{aligned}
\label{eq:four-level}
\end{equation}
with $\one$ being the identity operator. Here $H_0=\mathbf{p}^2/(2m)+V(\mathbf{r})$ is the state-independent part of the Hamiltonian that includes the kinetic energy of atomic motion in the $xy$ plane, as motion along the $z$ axis is frozen; $\mathbf r=(x,y)$ is a two-dimensional (2D) radius vector,
$\mathbf{p}=-\ii\hbar\boldsymbol\nabla$ is the corresponding 2D momentum operator, and $m$ is the atomic mass. As in previous studies ~\cite{LinMagnetic2009, Goldman2014}, we take the  amplitude of the Raman coupling $\Omega_R(x)$ to depend on one spatial coordinate  ($x$)
and the Raman phase $2k_Ry$ to  change in the orthogonal spatial direction  $y$. Thus $2\hbar k_R$ represents the Raman momentum transfer and $\omega_R=\hbar k_R^2/(2m)$ is the corresponding recoil frequency.
The auxiliary couplings $\omega_\lambda(x)$ are  considered to be
  $x$-dependent.

\subsection{Effective two-level description}
\label{sec:two-level}

For  sufficiently weak system-auxiliary coupling  $\omega_\lambda$ and a large enough auxiliary-state loss rate $\kappa_\lambda$,  one can eliminate the auxiliary-state manifold  ~\cite{Feshbach1958,Reiter2012}, leading to the effective Hamiltonian for the projected subspace of Raman-coupled states.  Specifically, we introduce
projectors onto the system of interest ($P$) and auxiliary doublets ($Q$),
\begin{equation}
 P=\sum_{\lambda\in\left\{\uparrow,\downarrow\right\}}
 \lvert\lambda\rangle\langle\lambda\rvert,\qquad
 Q=\sum_{\lambda\in\left\{\uparrow,\downarrow\right\}}
 \lvert a_\lambda\rangle\langle a_\lambda\rvert,
\label{eq:PQ-projectors}
\end{equation}
with $P+Q={\one}$. Projection of the stationary
Schr\"odinger equation onto the two subspaces gives
\begin{equation}
\begin{aligned}
 H_{P}(E)&=PH_4P+PH_4Q\,R(E)\,QH_4P,
\end{aligned}
\label{eq:projector-effective}
\end{equation}
where the resolvent is $R(E)=[E-QH_4Q]^{-1}$. We approximate the resolvent as $R(E)\approx
 -\frac{2\ii}{\hbar}\sum_\lambda\kappa^{-1}_\lambda|a_\lambda\rangle\langle a_\lambda|$.
Such an approximation is  applicable when the  loss rates for the auxiliary states greatly exceed the Raman coupling and the coupling to the auxiliary states,  $\kappa_\lambda \gg \Omega_R,\,|\omega_\lambda|$,  and $\hbar\kappa_\lambda$ is also much larger than the relevant motional energies. This is equivalent to the second-order Schrieffer-Wolff transformation~\cite{Bravyi2011,Starkov2023,Yanes22PRL}.

The resulting  projected Hamiltonian is

\begin{equation}
 H_2=
 H_0{\one}+V_{\mathrm{at-l}}(\mathbf{r})
 -\frac{\ii\hbar\Gamma(x)}{2}\one,
 \label{eq:H2-eff}
\end{equation}
where 
\begin{equation}
 V_{\mathrm{at-l}}(\mathbf{r})=\hbar\Omega_R(x)\ee^{\ii2k_Ry}\lvert\uparrow\rangle\langle\downarrow\rvert/2+\mathrm{H.c.}+\ii\frac{\hbar}{2}\Delta(x)\sigma_z
 \label{eq:V_at-l}
\end{equation}
is the atom-light coupling operator, $\Gamma(x)=\lvert\omega_\uparrow(x)\rvert^2/(2\kappa_\uparrow) + \lvert\omega_\downarrow(x)\rvert^2/(2\kappa_\downarrow)$ is the state-independent loss rate, $\Delta(x)=\lvert\omega_\downarrow(x)\rvert^2/(2\kappa_\downarrow) - \lvert\omega_\uparrow(x)\rvert^2/(2\kappa_\uparrow)$ is the effective detuning, and $\sigma_z=\lvert\uparrow\rangle\langle\uparrow\rvert-\lvert\downarrow\rangle\langle\downarrow\rvert$ is the Pauli $z$ operator.

\subsection{Adiabatic motion in the lowest dressed state}
\label{sec:projection}

 A dressed state 
$\lvert\chi^{\mathrm R}_s(\mathbf{r})\rangle$
is an eigenstate of the atom-light coupling operator given by  Eq.~\eqref{eq:V_at-l},
\begin{equation}
V_{\mathrm{at-l}} \lvert\chi^{\mathrm R}_s(\mathbf{r})\rangle = \varepsilon_s(\mathbf{r}) \lvert\chi^{\mathrm R}_s(\mathbf{r})\rangle\,,\label{eq:dressed-eigen}
\end{equation}
where $\varepsilon_s(\mathbf{r})$ is the corresponding eigenenergy, $\lvert\chi^{\mathrm R}_s(\mathbf{r})\rangle$ is the aforementioned (right) dressed state, and $s$  labels the eigenvalue~\cite{DalibardJuzeliunas2011}. For a non-Hermitian Hamiltonian, one needs to  use a biorthogonal eigenbasis~\cite{Brody2013}. We denote the left and right dressed states by
$\lvert\chi_s^{\mathrm L}\rangle$ and $\lvert\chi_s^{\mathrm R}\rangle$, respectively. For  the present two-component atom-light coupling  operator in Eq.~\eqref{eq:V_at-l}, 

 the eigenstate index  $s$ takes the values $s\in\{+,-\}$.  We choose the following normalization for biorthogonal basis vectors:
\begin{equation}
\langle\chi_s^{\mathrm R}\lvert\chi_s^{\mathrm R}\rangle=1,\qquad
 \langle\chi_s^{\mathrm L}\lvert\chi_{s'}^{\mathrm R}\rangle=\delta_{ss'}\,.
\label{eq:dressed-normalization}
\end{equation}
Such a convention is  useful for calculating experimental expectation values, which depend only on right states. 

Note that, in general, one cannot normalize simultaneously the three possible inner products $\langle \chi_s^{\mathrm L}|\chi_{s}^{\mathrm L}\rangle$, $\langle \chi_s^{\mathrm L}|\chi_{s}^{\mathrm R}\rangle$, and $\langle \chi_s^{\mathrm R}|\chi_{s}^{\mathrm R}\rangle$~\cite{MontagOzawaGeometry2026}.

If the energy gap between the dressed states $|\varepsilon_+(\mathbf{r})- \varepsilon_-(\mathbf{r})|$ is much larger than the  nonadiabatic couplings between the two dressed branches, the atoms will adiabatically follow a selected dressed state $s$.

In such an adiabatic approach, the full state vector of the Raman-coupled two-level system $\lvert\Psi_2(\mathbf{r},t)\rangle$ contains a contribution mainly from the selected dressed state: $\lvert\Psi_2(\mathbf{r},t)\rangle\approx\psi_s(\mathbf{r},t)\lvert\chi^{\mathrm R}_s(\mathbf{r})\rangle$, where $\psi_s(\mathbf{r},t)$ is a wavefunction of the atomic center of mass in the dressed state $\lvert\chi^{\mathrm R}_s(\mathbf{r})\rangle$. The dynamics of the dressed-state wavefunction is governed by the  adiabatic Hamiltonian (see Appendix \ref{app:adiabatic-proj})
\begin{equation}
H=\frac{1}{2m}\left(\mathbf{p}-\mathbf{A}_{s,s}(\mathbf{r})\right)^2+V(\mathbf{r})+W(\mathbf{r})+\varepsilon_s(\mathbf{r})-\frac{\ii\hbar\Gamma(\mathbf{r})}{2},\label{eq:projected-main}
\end{equation}
where $\mathbf A_{s,s^{\prime}}=i\hbar
\langle\chi_s^{\mathrm L}|\nabla\chi_{s^\prime}^{\mathrm R}\rangle$ and $W=\frac{1}{2m}\mathbf A_{+,-}\cdot\mathbf A_{-,+}$ are, respectively, the geometric vector and scalar potentials. 

In deriving Eq.~\eqref{eq:projected-main}, we projected the full state vector  $\lvert\Psi_2(\mathbf{r},t)\rangle$  onto the left state vector $\langle\chi_-^{\mathrm L}\lvert$, which is essential for the adiabatic
projection in non-Hermitian systems~\cite{MontagOzawaGeometry2026}.
Projection using the right-state vector $\langle\chi_-^{\rm R}\lvert$ would not give an equation for the adiabatic dynamics of atoms described by the wavefunction $\psi_-(\mathbf{r},t)$.

 The synthetic magnetic field corresponding to the vector potential $\mathbf{A}_{s,s}$ is given by $\mathbf{B}_s=\boldsymbol\nabla\times\mathbf{A}_{s,s}$.  Explicit expressions for the geometric vector and scalar potentials and the corresponding magnetic field for the present non-Hermitian Hamiltonian are presented in Appendix~\ref{app:adiabatic-proj}. 
 In the next subsection, we will consider the possibility of creating a constant imaginary magnetic field. 

\subsection{Engineering a uniform imaginary magnetic field}
\label{sec:uniform-field-eng}

We are interested in creating a uniform imaginary magnetic field  and a constant dressed-state gap using the imaginary detuning $\ii \Delta$ that is linear in the coordinate $x$,  like in previous studies of Hermitian systems~\cite{LinMagnetic2009,Goldman2014}. Additionally, we wish to minimize competing non-Hermitian terms, and thus we require the state-independent loss rate to be coordinate-independent.  This can be achieved by setting $\lvert\omega_{\uparrow,\downarrow}(x)\rvert^2=\kappa_{\uparrow,\downarrow}(\Gamma_0\mp\Omega_0\eta x)$, where we take $\Omega_0=\Omega_R(x=0)>0$. In that case, the imaginary part of the detuning and the overall loss read, respectively, $\Delta(x)=\Omega_0\eta x$ and  $\Gamma(x)=\Gamma_0$. Such a light beam profile could be realized with digital micromirror devices~\cite{Gauthier2016}.

Alternatively, such a two-level Hamiltonian can be achieved via two Gaussian light beams with center positions $x_{0,\lambda}$ and waists $w_\lambda$, i.e., $\omega_\lambda(x)=\sqrt{\kappa_\lambda\Gamma_0}\,
e^{-\left[(x-x_{0,\lambda})^2-x_{0,\lambda}^2\right]/w_\lambda^2}$. Overlapping displaced Gaussian beams have been utilized in optical trapping experiments~\cite{Shin2004}. If one chooses equal waists ($w_\lambda=w$) and opposite half-waist offsets ($x_{0,\lambda}=\mp w/2$), the mean loss is approximately uniform while the detuning is approximately linear in the coordinate $x$ near the origin $x=0$. In particular, one has $\Gamma(x)\approx \Gamma_0\left(1 - 4x^4/(3w^4)\right)\approx \Gamma_0$ and $\Delta(x)\approx\left(2\Gamma_0x/w\right)\left[1-4x^2/(3w^2)\right]\approx \Omega_0\eta x$, where now $\eta=2\Gamma_0/w\Omega_0$.

According to Eq.~\eqref{eq:Bplanar} of Appendix~\ref{app:adiabatic-proj}, the synthetic magnetic field is proportional to $\partial_x\left(\Delta(x)/\varepsilon_s(x)\right)$. For linear imaginary detuning $\ii \Delta(x)= \ii \Omega_0\eta x$, we therefore choose $\Omega_R(x)=\Omega_0\sqrt{1+\eta^2x^2}$, which  yields a constant dressed-state energy $\varepsilon_s=s\hbar\Omega_0/2$ and   provides a uniform imaginary magnetic field  $\mathbf{B}_s= -\ii s\Bcal\mathbf{e}_z$,  with $\Bcal=\hbar k_R\eta$\footnote{ The possibility of creating a constant (yet real) synthetic magnetic field was previously demonstrated for the Hermitian $\Lambda$ atom-light coupling scheme also using a proper atom-light coupling \cite{Juz05JPB}.}.  The constant Raman amplitude would instead produce a nonuniform  synthetic magnetic field and a gap closing at $|\Delta|=\Omega_R$, where the single-band adiabatic description breaks down. 
Furthermore, following the arguments presented above, we take the constant overall loss rate:  $\Gamma(x)=\Gamma_0$.

Nonnegative loss rates require $\Gamma_0\ge\Omega_0|\eta|L$ over $|x|\le L$, where $2L$ is the system size. Assuming that $|\eta x|\ll1$, we linearize the Cartesian component of the vector potential $A_x$ given by Eq.~\eqref{eq:A_x-Apend}, so that $A_x=-\ii\hbar\eta^2x/2$. This approximation yields a quadratic Hamiltonian, which can then be solved analytically. Furthermore, we compensate the geometric scalar potential $W$ given by Eq.~\eqref{eq:Wuniform} by choosing the external potential such that $V=-W$ in the populated region. We retain the square of this linearized vector potential in the resulting quadratic Hamiltonian.

Additionally, we omit the constant real term in the $y$ component of the vector potential $A_y$, which only shifts the momentum, as well as the constant dressed-state energy, which shifts the energy origin. Then, projecting the two-component Hamiltonian $H_2$ of Eq.~\eqref{eq:H2-eff} onto the lowest dressed state $|\chi_-^{\mathrm R}(\mathbf{r})\rangle$, we obtain the following effective Hamiltonian:

\begin{equation}
H= \frac{\left(p_x +\ii\hbar\eta^2x/2\right)^2+(p_y-\ii\mathcal{B}x)^2}{2m}-\frac{\ii\hbar\Gamma_0}{2},
\label{eq:Hcentral}
\end{equation}

where $\mathcal{B}=\hbar k_R\eta$ is the strength of the imaginary magnetic field.
Details of adiabatic projection are discussed in Appendix~\ref{app:adiabatic-proj}.  The Hamiltonian in Eq.~\eqref{eq:Hcentral} represents the central model, which we will study in the remainder of the paper.

\section{Gaussian dynamics in a uniform imaginary magnetic field}
\label{sec:gauss}

We wish to study the dynamics of a quantum particle in an imaginary magnetic field, governed by the Hamiltonian in Eq.~\eqref{eq:Hcentral}.  We shall consider a
Gaussian wavepacket with arbitrary initial position, mean momentum, and widths.
The projected state is
$|\Psi_2(\mathbf r,0)\rangle=\psi_0(\mathbf r)|\chi_-^{\mathrm R}(\mathbf r)\rangle$,
with envelope
\begin{equation}
\psi_0(\bm{r})=\mathcal N\exp\Big[\frac{\mathrm{i}\bm{p}_0\cdot\bm{r}}{\hbar}-\frac{(x-x_0)^2}{4\sigma_x^2}-\frac{(y-y_0)^2}{4\sigma_y^2}\Big]\,,
\label{eq:psi0}
\end{equation}
where $\mathcal N=(2\pi\sigma_x\sigma_y)^{-1/2}$ is the normalization factor,
$\mathbf r_0=(x_0,y_0)$ is the center position, and $\mathbf p_0=(p_{x0},p_{y0})$ is the mean momentum. We will also frequently use the wavenumbers, defined by $k_{i0}=p_{i0}/\hbar$.
The parameters $\sigma_x$ and $\sigma_y$ are the Gaussian spreads in both spatial directions. The Gaussian wavepacket can be approximately created by preparing
the ground state of a harmonic oscillator in one of the system spin
states, turning on real detuning and then slowly ramping the atom-light coupling and real detuning so that the
state adiabatically follows the lower dressed band, after which the trap
is released to initiate the dynamics governed by Eq.~\eqref{eq:Hcentral}.

In analogy to the real magnetic field, we define the cyclotron frequency
\begin{equation}
\omega=\mathcal B/m=\hbar k_R\eta/m.
\end{equation}

Note that one could understand the dynamics by studying the eigenspectrum. However, for non-self-adjoint differential operators, the eigenspectrum can be highly sensitive to perturbations, and one may need to study the pseudospectrum instead~\cite{TrefethenEmbree2005,Trefethen2005}.
While interesting, we do not pursue this question.

In the rest of this Section, we will first derive expressions for observables in terms of the dressed-state wavefunction. Secondly, we consider the non-Hermitian Ehrenfest equation for the first-order observables and their short-time dynamics. Thirdly, we will compare these predictions with the exact analytical solution. Finally, we shall study the position covariance $v_{ij}(t)$ and the associated squeezing dynamics.

It is noteworthy that wavepacket dynamics affected by non-Hermitian gauge fields was also considered in~\cite{MontagOzawa2026}.
Simulations have been carried out  in the symmetric gauge with a tight-binding lattice added~\cite{MontagOzawa2026}. On the other hand, we consider  continuum wavepacket dynamics in the Landau-type gauge with an extra Cartesian component $A_x$  of the vector potential given by Eq.~\eqref{eq:A_x-Apend}, which appears after projecting the exact dynamics onto one of the dressed states.
The two gauges are related by a nonunitary transformation which changes the wavefunction amplitude. In particular, the same Gaussian profiles are not equivalent  in both setups. Furthermore, the paper~\cite{MontagOzawa2026} considered a situation where the real part of the magnetic field is larger than the imaginary part, so the cyclotron motion plays an important role in the dynamics. On the other hand, we are considering another regime in which the dynamics is affected by a purely imaginary magnetic field.

\subsection{Physical observables}
\label{sec:conditional-observables}

Let us consider nuances of calculating observables for our system. It is simplest to start from the two-component description and then derive expressions for the adiabatically projected model.

For the two-component state $|\Psi_2(\mathbf{r}, t)\rangle$, the Born rule gives the surviving fraction of atoms and the observable averages,
\begin{align}
N(t)&=\int\dd^2r\,\langle\Psi_2|\Psi_2\rangle,\notag\\
\langle O\rangle_t&=\frac{1}{N(t)}\int\dd^2r\,
\langle\Psi_2|O|\Psi_2\rangle.
\label{eq:spinor-observables}
\end{align}
Within the adiabatic approximation, $|\Psi_2(\mathbf{r}, t)\rangle=\psi|\chi_-^{\rm R}\rangle$. Since we have chosen the normalization $\langle\chi_-^{\mathrm R}|\chi_-^{\mathrm R}\rangle=1$,  the surviving fraction and the position moments are given by
\begin{align}
N(t)&=\int\dd^2r\,|\psi|^2,\notag\\
\langle r_i^mr_j^n\rangle_t&=\frac{1}{N(t)}\int\dd^2r\,
r_i^mr_j^n |\psi|^2,
\label{eq:x-observables}
\end{align}
where $i,j\in\{x,y\}$ and $m,n\in\mathbb{N}$.

For momentum, differentiation also acts on the dressed state; however, we have $\langle\chi_-^{\mathrm R}|\nabla\chi_-^{\mathrm R}\rangle=0$ in the chosen gauge, leading to
\begin{align}
  \langle\mathbf p\rangle_t
  &=\frac{1}{N(t)}\int\dd^2r\, \langle\Psi_2|(-i\hbar\nabla)|\Psi_2\rangle=\\
  &=-\frac{i\hbar}{N(t)}\int\dd^2r\, \left[\psi^*\nabla\psi + |\psi|^2\langle\chi_-^{\mathrm R}|\nabla\chi_-^{\mathrm R}\rangle\right]=\\
  &=\frac{1}{N(t)}\int\dd^2r\, \psi^*(-i\hbar\nabla)\psi.
\end{align}
Since we have no vector potential at the level of the two-component system, the kinetic momentum coincides with the canonical momentum. More surprisingly, the mean physical momentum also equals the mean canonical momentum calculated from the projected wavefunction, even though we have a nonzero imaginary magnetic field. This result is in contrast to the real gauge field case, where the kinetic and canonical momenta do not coincide.

Thus, the first-order position and momentum averages reduce to
\begin{equation}
\langle O\rangle_t=\frac{1}{N(t)}\int\dd^2r\,\psi^*O\psi.
\label{eq:observables}
\end{equation}
Higher-order observables involving products of momenta generally do not follow this simple reduction and acquire correction terms containing spatial derivatives of the dressed state. Notice that observables involve only the right dressed states. The left dressed states enter only the projection that determines the effective Hamiltonian.

These averages describe measurements on the surviving atoms and will be studied in the following sections.

\subsection{The non-Hermitian Ehrenfest equation and short-time dynamics}

Now we will study the short-time dynamics of observables via the non-Hermitian Ehrenfest equation~\cite{TaoSpielman2026,GraefeSchubert2011}. We write the first moments as $x(t)$, $y(t)$, $p_x(t)$, and $p_y(t)$,
and define the position covariance by
\begin{equation}
 v_{ij}(t)=\langle r_i r_j\rangle_t
 -\langle r_i\rangle_t\langle r_j\rangle_t,\qquad i,j\in\{x,y\}.
 \label{eq:covdef}
\end{equation}

The Ehrenfest equation for non-Hermitian systems is given by
\begin{equation}
\begin{aligned}
\frac{\dd\langle O\rangle_t}{\dd t}
&=\frac{\ii}{\hbar}\langle[H_{\rm h},O]\rangle_t
+\frac{1}{\hbar}\langle\{H_{\rm ah},O\}\rangle_t\\
&\quad-\frac{2}{\hbar}\langle H_{\rm ah}\rangle_t\langle O\rangle_t,
\end{aligned}
\label{eq:ehr}
\end{equation}
where $H=H_{\rm h}+\ii H_{\rm ah}$, with both $H_{\rm h}$ and
$H_{\rm ah}$ Hermitian. Appendix~\ref{app:ehr} gives the derivation of this equation.

For the imaginary magnetic field Hamiltonian in Eq.~\eqref{eq:Hcentral}, one has
\begin{align}
H_{\rm h}&=\frac{\mathbf p^2}{2m}-\frac{m\omega^2x^2}{2}-\frac{\hbar^2\eta^4x^2}{8m}
,\label{eq:Hh}\\
H_{\rm ah}&=\frac{\hbar\eta^2}{4m}\{x,p_x\}
-\omega xp_y-\frac{\hbar\Gamma_0}{2}.\label{eq:Hah}
\end{align}

The Hermitian part $H_{\rm h}$ describes an inverted harmonic oscillator. The anti-Hermitian part $\ii H_{\rm ah}$ contains homogeneous loss proportional to $\Gamma_0$ and nonhomogeneous terms coming from the vector potential $\mathbf A$. The $-\ii\omega xp_y$ term in the Hamiltonian contributes $-2\omega\langle xp_y\rangle_t/\hbar$ to the relative rate of change of the surviving fraction, $\dot N/N$, leading to different position and momentum components being attenuated at different rates. Such nonuniform loss can give rise to exotic motion like self-acceleration.

Let us consider a special initial condition in Eq.~\eqref{eq:psi0}, which corresponds to a centered, initially circular packet launched along the Raman direction,
\begin{equation}
 x_0=y_0=p_{x0}=0,\qquad p_{y0}\ne0,\qquad
 \sigma_x=\sigma_y=\sigma.
 \label{eq:y-launch}
\end{equation}
Using the Ehrenfest equation of Eq.~\eqref{eq:ehr}, the short-time dynamics for the position is given by
\begin{align}
x(t)&=-\frac{2\omega\sigma^2p_{y0}}{\hbar}\,t+O(t^3)\label{eq:y-launch-drift_x},\\
y(t)&=\frac{p_{y0}}{m}\,t+O(t^3)\label{eq:y-launch-drift_y}.
\end{align}

The vector-potential cross term $-\ii\omega xp_y$ in the Hamiltonian attenuates the two sides of the moving packet at different rates. Over time, the side with $\omega xp_{y0}<0$ experiences less loss, leading to center-of-mass motion with velocity proportional to $p_{y0}$. Additionally, the anticommutator terms in the Ehrenfest equation lead to width-dependent dynamics of the first position moments. For arbitrary initial states, anti-Hermitian terms couple $n$th-order moments to $\left(n+2\right)$th-order moments. However, for the case of the Gaussian wavepacket, the higher-order moments simplify into first- and second-order moments. In particular, broader wavepackets experience larger transverse drift. The following dynamics is different from the real magnetic field case, where the first-order moments do not couple to higher moments.

\begin{figure}[t]
 \centering
  \includegraphics[width=\columnwidth]
   {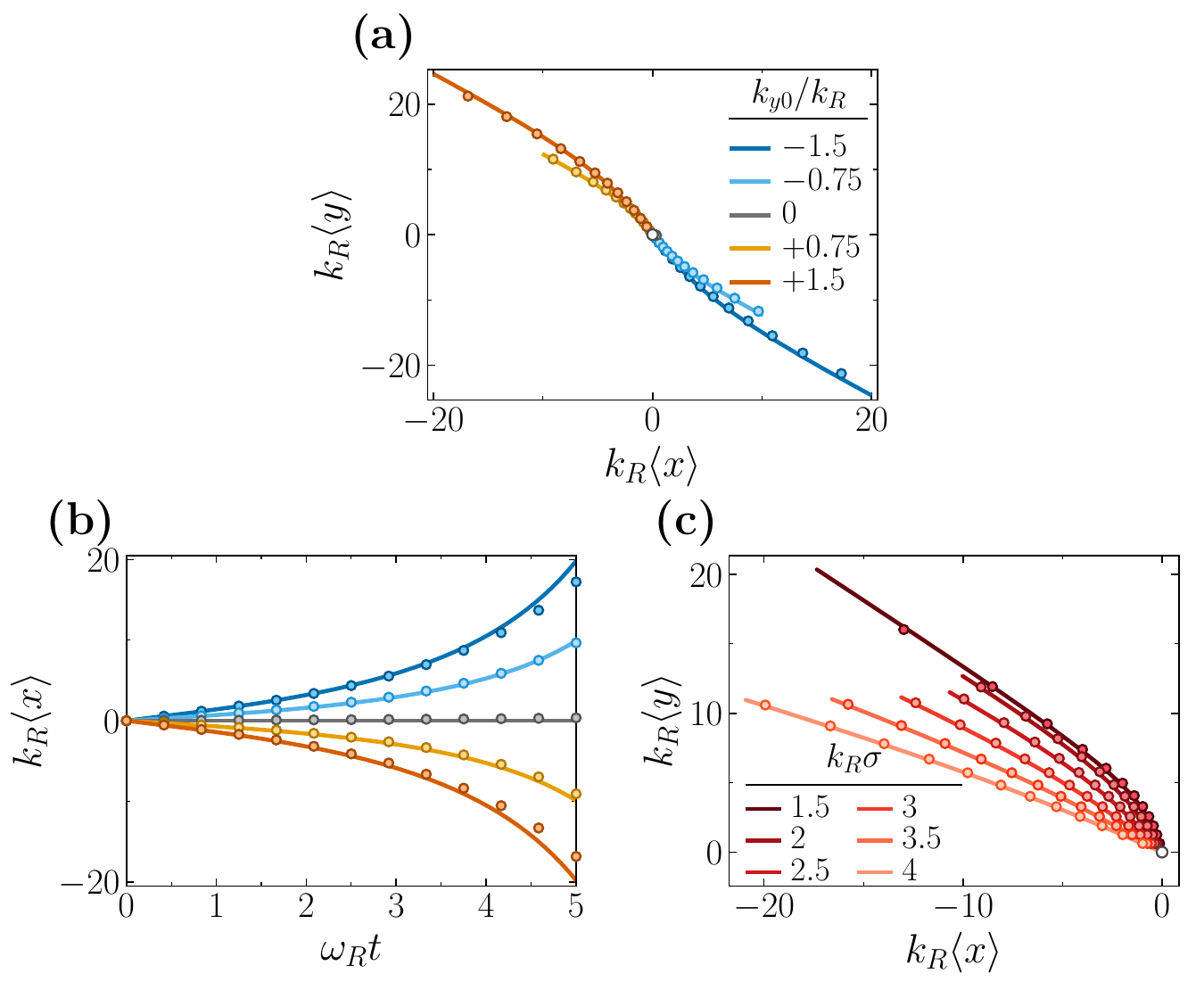}
   \caption{Center-of-mass dynamics. (a) Trajectories of an initially centered Gaussian wavepacket for a range of launch momenta $k_{y0}/k_R$. (b) Corresponding $x$ center-of-mass evolution, with fixed width $k_R\sigma=k_R\sigma_x=k_R\sigma_y=3/\sqrt{2}$. (c) Trajectories of an initially centered Gaussian wavepacket for a range of different initial widths $k_R\sigma$, with fixed initial momentum $k_{y0}/k_R=0.75$. Markers are exact numerical results of the projected model using the precise form of $A_x$ given by Eq.~\eqref{eq:A_x-Apend}, and solid curves are analytical results of the projected model, obtained by linearizing $A_x$. Here $k_{x0}/k_R=0$ and $\eta/k_R=0.05$.}
   \label{fig:CM-dynamics}
\end{figure}

Analyzing short-time dynamics for the momenta, one has
\begin{align}
p_x(t)
&=-\omega p_{y0}t^2
\Biggl[\frac{2m\omega^2\sigma^2}{\hbar}
\notag\\
&\quad+\frac{\hbar(1+\eta^2\sigma^2)(1+2\eta^2\sigma^2)}
{4m\sigma^2}\Biggr]+O(t^4),\\
p_y(t)
&=p_{y0}\left(1+\omega^2t^2\right)+O(t^4).
\end{align}
We see that the transverse momentum $p_x(t)$ grows quadratically in time due to the $xp_y$ term in the Hamiltonian. The magnitude of growth is governed by both position and momentum uncertainties. Terms proportional to $\sigma^2$ describe initial position uncertainty, whereas terms proportional to $1/\sigma^2$ describe initial momentum uncertainty via the Heisenberg uncertainty principle.

Although $[H,p_y]=0$, the longitudinal momentum $p_y(t)$
changes because different momentum components decay at different rates.
Thus, the initial transverse drift precedes
momentum deflection, while the longitudinal acceleration results from
momentum-selective loss.

At fixed initial width, both displacements are proportional to the launch momentum
$p_{y0}$. Reversing the launch direction, i.e., $p_{y0} \rightarrow -p_{y0}$, therefore reflects the entire trajectory
through the origin, while increasing its magnitude increases the
spatial excursion, as illustrated in
Fig.~\hyperref[fig:CM-dynamics]{\ref*{fig:CM-dynamics}(a)} and Fig.~\hyperref[fig:CM-dynamics]{\ref*{fig:CM-dynamics}(b)}.

The width dependence in Fig.~\hyperref[fig:CM-dynamics]{\ref*{fig:CM-dynamics}(c)} reflects the
coupling between the center-of-mass position and the wavepacket spread. Over the
range shown, broader packets undergo a larger transverse displacement along $x$ at a given longitudinal position. They sample a greater variation in
attenuation across $x$, producing a stronger redistribution of the
surviving density.

\subsection{Exact analytical solution}
The quadratic projected Hamiltonian in Eq.~\eqref{eq:Hcentral} preserves Gaussian wavepackets and is
analytically solvable. A nonunitary transformation to eliminate $A_x$, followed by an imaginary
position shift to eliminate $A_y$, reduces each conserved $p_y$ sector to an inverted harmonic oscillator. Then one may evolve the wavefunction via the inverted-oscillator propagator and transform back to obtain the dynamics of the dressed-state wavefunction.
Once the wavefunction is found, we calculate the first moments, the position covariances, and the surviving fraction of atoms. The detailed derivation is given in Appendix~\ref{app:gaussian}.

\begin{figure*}[t]
 \centering
 \includegraphics[width=\textwidth]{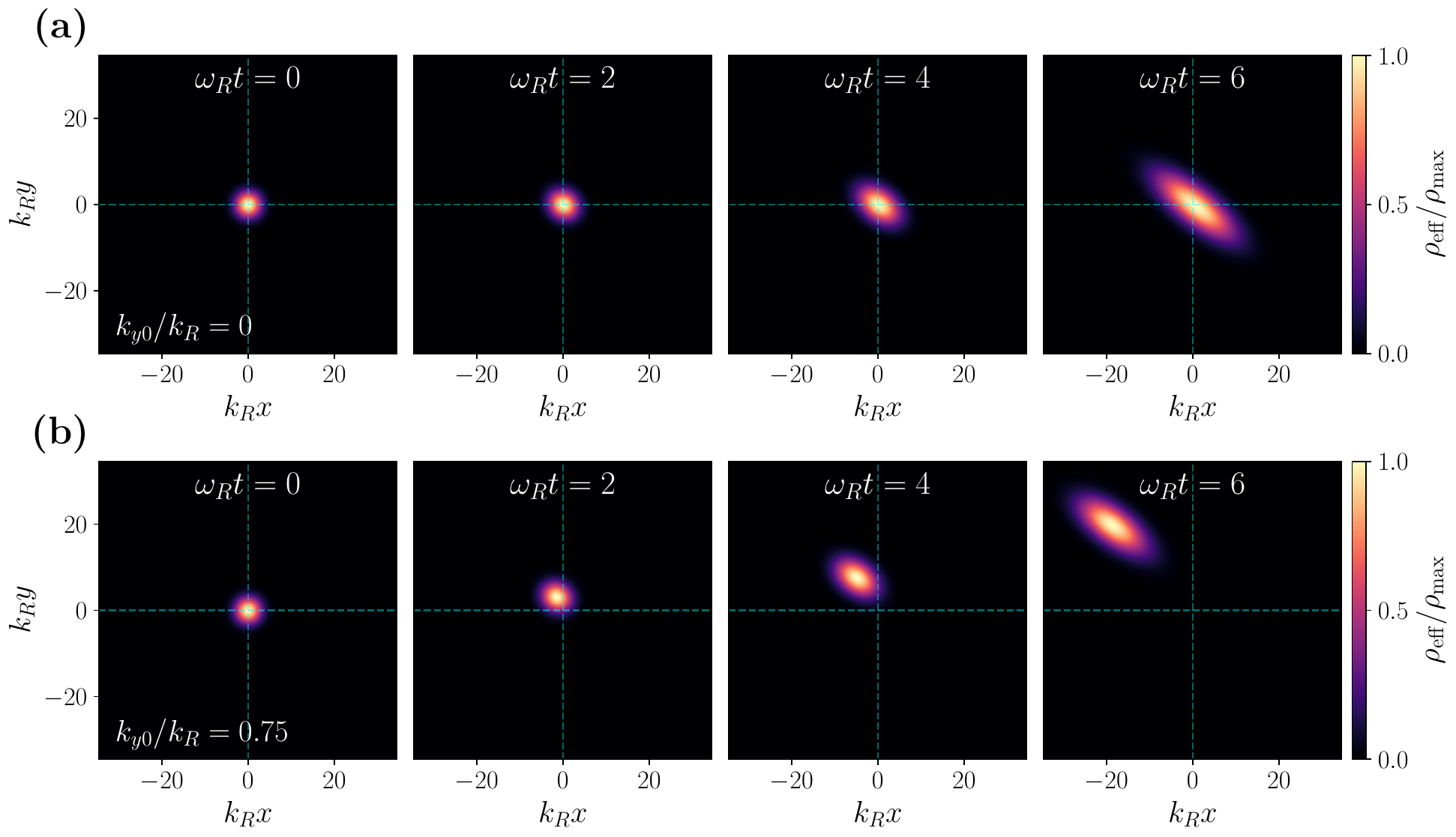}
 \caption{Density evolution. Results are obtained by exact numerics of the projected model using the precise form of $A_x$ given by Eq.~\eqref{eq:A_x-Apend}.
(a) Density profiles for an initially centered Gaussian wavepacket with
$k_{x0}=0$ and $k_{y0}=0$.
(b) Same as in (a), but for $k_{y0}=0.75\,k_R$.
Snapshots are shown at the times indicated in the panels.
Here $k_R\sigma_x=k_R\sigma_y=3/\sqrt{2}$ and
$\eta/k_R=0.05$.
Each density is normalized to its instantaneous maximum,
$\rho_{\max}(t)$, so the color scale displays the spatial evolution of
the wavepacket rather than changes in its absolute population.}
 \label{fig:sigma3-density-snapshots}
\end{figure*}

For the centered launch along $y$, i.e.,
$x_0=y_0=p_{x0}=0$, $p_{y0}\ne0$, and $\sigma_x=\sigma_y=\sigma$,
the four first moments reduce to
\begin{subequations}\label{eq:exact-launch}
\begin{align}
x(t)&=-\frac{2p_{y0}s}{\hbar\mathcal D_\eta}
\left[\sigma^2c+(1+\epsilon)^2
\left(\frac{\hbar}{2m\omega\sigma}\right)^2(c-1)\right]\,,
\label{eq:exact-x}\\[2pt]
y(t)&=\frac{p_{y0}s}{m\omega\mathcal D_\eta}
\left\{1-[\epsilon+2\delta(1+\epsilon)](c-1)\right\}\,,
\label{eq:exact-y}\\[2pt]
p_x(t)={}&-\frac{2m\omega p_{y0}}{\hbar\mathcal D_\eta}
\Biggl[(1+\delta)\sigma^2s^2
\notag\\
&\quad +(1+\epsilon)
\left(\frac{\hbar}{2m\omega\sigma}\right)^2
c(c-1)\Biggr],
\label{eq:exact-px}\\[2pt]
p_y(t)&=\frac{D_x}{\mathcal D_\eta}p_{y0}\,.
\label{eq:exact-py}
\end{align}
\end{subequations}
Here $c=\cosh(\omega t)$, $s=\sinh(\omega t)$,
$\epsilon=\eta^2\sigma^2$, and
$\delta=(1+\epsilon)/(4k_R^2\sigma^2)$, with
\begin{align}
D_x&= 1-[\epsilon+\delta(1+\epsilon)]s^2\,,\notag\\
\mathcal D_\eta&= D_x-s^2-(1+\epsilon)
\left(\frac{\hbar}{2m\omega\sigma^2}\right)^2(c-1)^2\,.
\label{eq:exact-denominators}
\end{align}
For our considered time interval, we have $\mathcal D_\eta>0$. Eventually, $\mathcal D_\eta$ reaches zero, indicating the breakdown of the quadratic description.

The exact analytical solutions describe dynamics beyond the short-time expansion, e.g., the departure from the initial
straight trajectory: the transverse displacement grows increasingly
rapidly relative to the longitudinal displacement as the packet evolves.

Expanding Eq.~\eqref{eq:exact-launch} reproduces the short-time results:
$x$ and $y$ grow linearly in time, while $p_x$ and $p_y-p_{y0}$ grow quadratically.
The transverse density shift therefore precedes the buildup of transverse
momentum.

All the aforementioned dynamics are also observed in exact numerics, shown in Fig.~\ref{fig:sigma3-density-snapshots}. Even for zero launch momentum, the packet develops an
inclined density profile while its center remains at the origin,
as seen in Fig.~\hyperref[fig:sigma3-density-snapshots]{\ref*{fig:sigma3-density-snapshots}(a)}. A finite positive
launch momentum additionally produces propagation along $y$ and a
transverse displacement toward negative $x$, evident in
Fig.~\hyperref[fig:sigma3-density-snapshots]{\ref*{fig:sigma3-density-snapshots}(b)}. The density profiles thus
reveal both the displacement of the surviving cloud and its deformation
about the moving center.

\subsection{Position correlations and squeezing of the launched packet}
\label{sec:position-squeezing}

We now study how the packet spreads about its moving center. The
position covariance matrix is
\begin{equation}
\mathbf v(t)=\begin{pmatrix}v_{xx}(t)&v_{xy}(t)\\
v_{xy}(t)&v_{yy}(t)\end{pmatrix}\,,
\label{eq:position-covariance-matrix}
\end{equation}

where $v_{ij}(t)$ is given by Eq.~\eqref{eq:covdef}. The diagonal entries are the position variances, with standard deviations
$\sigma_x(t)=\sqrt{v_{xx}(t)}$ and $\sigma_y(t)=\sqrt{v_{yy}(t)}$.
The off-diagonal entry $v_{xy}$ describes correlations between the two
coordinates. For the initially circular packet of Eq.~\eqref{eq:y-launch},
$\mathbf v(0)=\sigma^2\one_2$.

Expanding the exact covariance in Appendix~\ref{app:gaussian-covariances}
gives
\begin{subequations}\label{eq:y-launch-variances}
\begin{align}
v_{xx}(t)&=\sigma^2+\left[(2+\epsilon)\omega^2\sigma^2
+\frac{\hbar^2(1+\epsilon)^3}{4m^2\sigma^2}\right]t^2
\notag\\
&\quad+O(t^4),\\
v_{yy}(t)&=\sigma^2+\frac{\hbar^2}{4m^2\sigma^2}t^2+O(t^4),\\
v_{xy}(t)&=-\frac{\hbar\omega}{4m}(3+\epsilon)t^2+O(t^4),
\label{eq:y-launch-correlation}
\end{align}
\end{subequations}
where $\epsilon=\eta^2\sigma^2$. Both laboratory variances initially
increase. The terms proportional to $\sigma^2$ and $1/\sigma^2$
again describe the contributions of initial position and momentum
uncertainty. These expressions depend on the initial spreads, but not
on the first moments. This independence also holds for the exact
covariance with unequal initial widths. Thus, changing the launch
momentum changes the center-of-mass motion and surviving population,
while leaving the normalized shape unchanged. Uniform loss likewise
cancels from the covariance.

For the separable initial state considered here, the cross-covariance
$v_{xy}$ is generated by the imaginary magnetic field and vanishes
in its absence. This correlation tilts the density ellipse away from
the laboratory axes. Its orientation generally changes with time,
giving a rotation of the squeezing axes. Reversing the field changes
the sign of $v_{xy}$ and reflects these axes, while leaving the
principal spreads unchanged.

The spreads along the principal axes follow from the eigenvalues
$\lambda_+\ge\lambda_-$ of $\mathbf v$,
\begin{equation}
\lambda_\pm=\frac{v_{xx}+v_{yy}}{2}
\pm\frac12\sqrt{(v_{xx}-v_{yy})^2+4v_{xy}^2}.
\label{eq:main-principal-variances}
\end{equation}
The corresponding orthogonal eigenvectors are
\begin{equation}
\begin{aligned}
\mathbf e_+&=(\cos\theta_+,\sin\theta_+)^T,\\
\mathbf e_-&=(-\sin\theta_+,\cos\theta_+)^T,\\
\theta_+&=\tfrac12\operatorname{atan2}(2v_{xy},v_{xx}-v_{yy}).
\end{aligned}
\label{eq:main-principal-directions}
\end{equation}
The maximum spread is $\sqrt{\lambda_+}$ along $\mathbf e_+$, and
the minimum spread is $\sqrt{\lambda_-}$ along $\mathbf e_-$.
Initially the two eigenvalues coincide, so the circular density has
no preferred direction. The directions become defined as the packet
deforms.

Although $v_{xx}$ and $v_{yy}$ both grow, the correlations can reduce
$\lambda_-$ below its initial value. Substitution of
Eq.~\eqref{eq:y-launch-variances} into
Eq.~\eqref{eq:main-principal-variances} shows that one principal
spread initially increases while the other decreases when
\begin{equation}
|\omega|>\frac{\hbar}{m\sigma^2}\sqrt{1+\eta^2\sigma^2}.
\label{eq:y-launch-narrowing}
\end{equation}
Below this threshold, both principal spreads initially increase.
At equality, the exact solution gives $\lambda_-(t)=\sigma^2$ and
fixed principal directions throughout the normalizable evolution,
while $\lambda_+(t)$ increases. Appendix~\ref{app:circular-launch}
gives the derivation and the finite-time expressions. Thus the
eigenvalues determine the broadening and narrowing, and the eigenvectors
give the orientation of the squeezed packet.

The origin of these effects can be understood from the transformed
Hamiltonian. The inverted harmonic oscillator
$p_x^2/(2m)-m\omega^2x^2/2$ generates squeezing of conjugate
quadratures in the $x,p_x$ phase plane.
On its own, it does not couple $x$ and $y$ and therefore cannot rotate
the spatial density ellipse of an initially separable packet.
The imaginary-field coupling $-\ii\omega xp_y$, together with
kinetic propagation, generates $v_{xy}$ and the rotation of the
spatial squeezing axes. It also changes the principal spreads, so
the spatial narrowing condition in Eq.~\eqref{eq:y-launch-narrowing}
is a property of the full dynamics. The resulting spreads and their
orientation can be measured directly from the density profiles.

\section{Imaginary Aharonov-Bohm transport}
\label{sec:sig}

An imaginary magnetic flux changes the amplitude acquired during
transport around a closed path. We study this effect by moving a
localized trapped mode through a uniform imaginary field and comparing
the surviving populations for opposite orientations. The ratio of these populations depends on the enclosed area, providing a geometric analogue of the Aharonov-Bohm (AB) effect.

\begin{figure}[htb]
 \centering
  \includegraphics[width=\columnwidth]
   {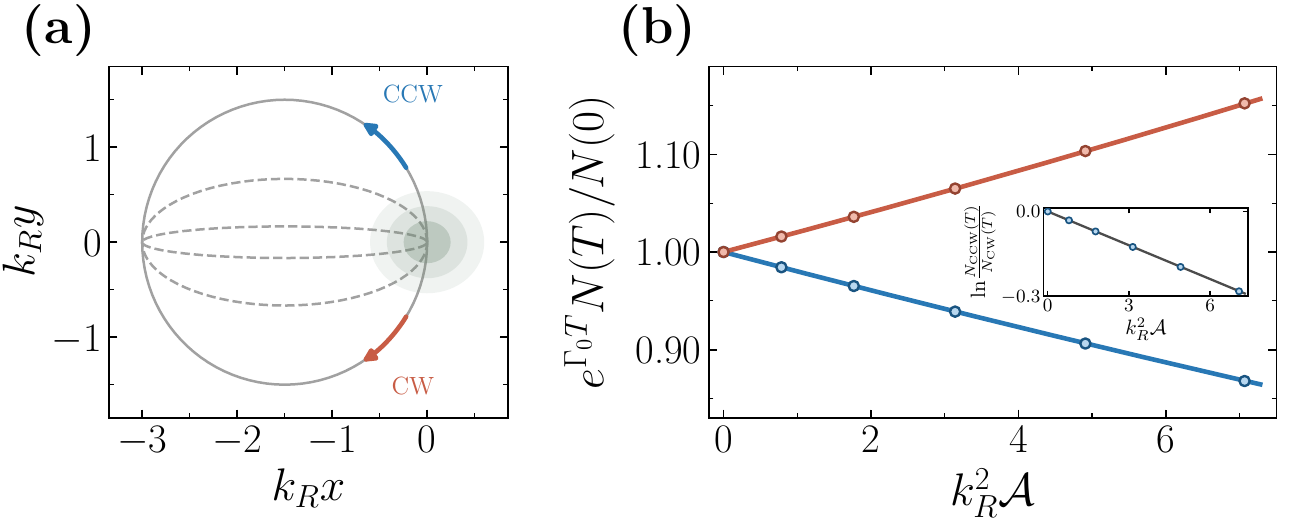}
     \caption{Imaginary AB transport.
    (a) Localized trapped mode transported along closed elliptical paths
    in opposite orientations, indicated by the arrows. Dashed curves show
    paths with smaller enclosed areas.
    (b) Background-loss-removed norm, $e^{\Gamma_0T}N(T)/N(0)$, after one
    cycle as a function of the unsigned enclosed area $\mathcal A$ for
    clockwise and counterclockwise transport. Markers show finite-time
    numerical evolution retaining the full $A_x$, while solid curves show
    the adiabatic quadratic prediction. The inset shows
    $\ln[N_{\rm CCW}(T)/N_{\rm CW}(T)]$, with the prediction of
    Eq.~\eqref{eq:ab-uniform-ratio}. Here $T\omega_R=20$, $\eta/k_R=0.01$,
    $(\Omega_x,\Omega_y)/\omega_R=(8,10)$, $k_RR_x=1.5$, and $R_y$
    is varied to change the enclosed area.}
    \label{fig:AB}
\end{figure}

We add an anisotropic harmonic trap to Eq.~\eqref{eq:Hcentral},
\begin{equation}
V_{\rm tr}(t)=\frac m2\left[\Omega_x^2(x-X)^2+
\Omega_y^2(y-Y)^2\right],
\label{eq:ab-trap}
\end{equation}
where $\mathbf R=(X,Y)$ is the trap center. At $\mathbf R=0$, we
prepare the stationary right mode $\phi_0$ that is continuously
connected to the harmonic oscillator ground state as the gauge coupling
vanishes. This mode is normalizable.

The trap follows an ellipse with a
linear angular ramp,
\begin{equation}
\begin{aligned}
\theta(t)&=2\pi t/T,\qquad 0\le t\le T,\\
X(t)&=R_x[\cos\theta(t)-1],\qquad
Y_\pm(t)=\pm R_y\sin\theta(t).
\end{aligned}
\label{eq:ab-ellipse}
\end{equation}

For $R_x,R_y>0$, the $+$ path is counterclockwise (CCW) and the $-$
path is clockwise (CW). Both paths begin and end at the origin and enclose areas $\mathcal A_\pm=\oint X\,\mathrm dY_\pm=\pm\pi R_xR_y$. Fig.~\ref{fig:AB} uses the unsigned area $\mathcal{A}=\pi R_x R_y$.

For the quadratic model of Eq.~\eqref{eq:Hcentral}, translation of the trap by $\mathbf R$ gives the unnormalized stationary mode
\begin{equation}
\phi_{\mathbf R}(\mathbf r)
=
\ee^{\eta^2X(x-X)/2-\mathcal B X(y-Y)/\hbar}
\phi_0(\mathbf r-\mathbf R).
\label{eq:ABphi}
\end{equation}
The exponential factor compensates the change in the vector potential
under translation. The eigenvalue is therefore independent of $\mathbf{R}$ and equals $E_0-\ii\hbar\Gamma_0/2$, where $E_0$ is the real
energy of the centered mode.
In the slow-transport limit, the state follows this mode.

We introduce the corresponding left mode, normalized by $\langle\widetilde\phi_{\mathbf R}|\phi_{\mathbf R}\rangle=1$. Since the centered modes have even parity, the terms odd under spatial inversion do not
contribute to the overlap, giving
\begin{equation}
\langle\widetilde\phi_{\mathbf R}|\partial_t\phi_{\mathbf R}\rangle
=-\frac{\eta^2}{2}X\dot X+\frac{\mathcal B}{\hbar}X\dot Y.
\label{eq:ab-connection}
\end{equation}

In the slow-transport limit, we write $\psi\simeq c(t)\phi_{\mathbf R(t)}$. Projecting onto the
left mode gives
$\dot c/c=-\ii E_0/\hbar-\Gamma_0/2
-\langle\widetilde\phi_{\mathbf R}|\partial_t\phi_{\mathbf R}\rangle$
\cite{GarrisonWright1988,OzawaSchomerus2025}.
For a closed path, $\oint X\,\dd X=0$, so $A_x$ makes no net
geometric contribution. The stationary mode and its norm return to
their initial values. The geometric amplitude factor for the positively
oriented path and the populations for the two orientations are
therefore
\begin{align}
W_C&=\exp\left(\frac{\ii}{\hbar}\oint_C\mathbf A\cdot\dd\mathbf r\right)
=\ee^{-\mathcal B\mathcal A/\hbar},\label{eq:AB}\\
N_\pm(T)&\simeq\ee^{-\Gamma_0T}
\ee^{\mp2\mathcal B \mathcal{A}/\hbar},\label{eq:ab-uniform-norm}\\
\ln\frac{N_+(T)}{N_-(T)}&\simeq-\frac{4\mathcal B \mathcal{A}}{\hbar}.
\label{eq:ab-uniform-ratio}
\end{align}
Here $N=\langle\psi|\psi\rangle$ is the squared norm, with
$N(0)=1$. In the adiabatic limit, the geometric amplitude factor depends only on the signed area $\mathcal{A}_{\pm}$ enclosed by the path. Since the atoms move through a nonzero field, this response is a
geometric AB analogue~\cite{OzawaHayata2024}. For the field orientation in
Fig.~\hyperref[fig:AB]{\ref*{fig:AB}(b)}, CW transport enhances
the population relative to the uniform background decay, while
CCW transport reduces it.

The linear ramp gives a nonzero initial trap velocity, which
can produce deviations from the adiabatic result. Nevertheless, the logarithmic
population ratio obtained numerically from Eq.~\eqref{eq:Hfinal} differs
from the prediction of Eq.~\eqref{eq:ab-uniform-ratio} by less than
$0.26\%$ over the nonzero areas shown in the inset of
Fig.~\hyperref[fig:AB]{\ref*{fig:AB}(b)}.

\section{Validity of the adiabatic model}

\label{sec:validity}

\begin{figure}[htb]
 \centering
 \includegraphics[width=\columnwidth]{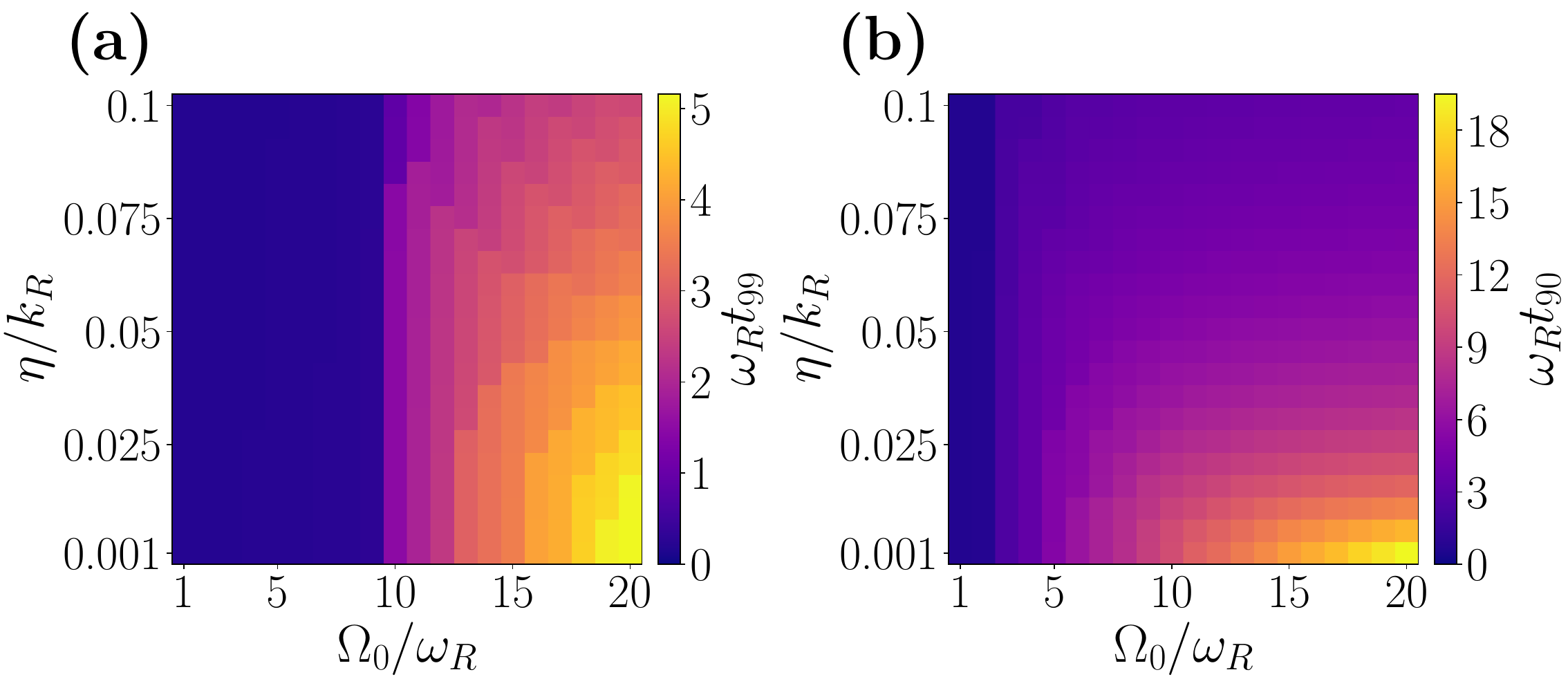}
 \caption{Validity of the adiabatic description. Color shows the dimensionless
    times (a) $\omega_Rt_{99}$ and (b) $\omega_Rt_{90}$, where $t_{99}$
    and $t_{90}$ are the first times at which the normalized fidelity
    between the two-level state and the reconstructed effective spinor
    falls to $0.99$ and $0.9$, respectively. The initial Gaussian
    envelope is centered, with $k_{x0}=k_{y0}=0$ and
    $k_R\sigma_x=k_R\sigma_y=3/\sqrt{2}$.}
 \label{fig:adiabatic-fidelity-diagram}
\end{figure}

In this section, we test the adiabatic description by comparing evolution under the
two-level Hamiltonian $H_2$ with $V=-W$ given by Eq.~\eqref{eq:H2-eff} with the
projected dynamics governed by Eq.~\eqref{eq:Hfinal}. The projected
Hamiltonian retains the full spatial dependence of the geometric vector
potential, allowing us to test the single-band approximation
independently of the spatial linearization used in
Eq.~\eqref{eq:Hcentral}.

In both calculations we use the same initial state, $\psi_0(\mathbf r)|\chi_-^{\mathrm R}(\mathbf r)\rangle$, where $\psi_0$ is the Gaussian envelope given by Eq.~\eqref{eq:psi0}. After evolving the projected envelope, we reconstruct the effective spinor $|\Psi_{\rm eff}(\mathbf r,t)\rangle=\psi(\mathbf r,t)|\chi_-^{\mathrm R}(\mathbf r)\rangle$. Its agreement with the full two-level state $|\Psi_{\rm full}(\mathbf r,t)\rangle=|\Psi_2(\mathbf r,t)\rangle$ is measured by the normalized fidelity
\begin{equation}
F(t)=
\frac{|\langle\Psi_{\rm full}(t)|\Psi_{\rm eff}(t)\rangle|^2}
{\langle\Psi_{\rm full}(t)|\Psi_{\rm full}(t)\rangle
 \langle\Psi_{\rm eff}(t)|\Psi_{\rm eff}(t)\rangle}.
\label{eq:numerical-spinor-fidelity}
\end{equation}

The normalized fidelity is independent of uniform background loss. We define $t_f$ as the first time at which the fidelity falls to a
specified threshold, $F(t_f)=f$. This time characterizes the interval
over which the projected description remains accurate for the
surviving atoms at the chosen fidelity threshold.

Fig.~\ref{fig:adiabatic-fidelity-diagram} shows the crossing times
$t_{99}$ and $t_{90}$ for thresholds $0.99$ and $0.9$, respectively.
For the centered Gaussian preparation considered, increasing
$\Omega_0/\omega_R$ extends the interval of agreement, whereas
increasing $\eta/k_R$ shortens it. The $t_{99}$ map depends primarily
on the Raman amplitude $\Omega_0$, while the longer $t_{90}$ window
shows a stronger dependence on the field strength. The validity window at the chosen fidelity threshold can therefore be extended by increasing the Raman coupling amplitude.

\section{Conclusions}
\label{sec:conclusions}

We have proposed a dissipative Raman scheme for generating a uniform
imaginary magnetic field for ultracold atoms. Complementary removal
channels produce a linear differential detuning and uniform background
loss, while a spatially varying Raman coupling maintains a constant
dressed-state gap. The adiabatic projection gives an imaginary vector
potential with nonzero curl. 

For the quadratic projected Hamiltonian, nonunitary transformations
reduce the dynamics to an inverted harmonic oscillator and yield
exact Gaussian first moments, covariances, and surviving populations.
For a packet launched along the direction of Raman recoil,
attenuation produces a transverse density shift linear in time,
preceding the quadratic onset of transverse momentum. The covariance
depends on the initial spreads independently of the first moments.
The inverted oscillator generates quadrature squeezing, while the
imaginary-field coupling produces spatial correlations and rotation
of the principal axes. Above the derived threshold, one principal
position spread initially narrows, while the other broadens.

Cyclic transport in a moving trap gives an imaginary Aharonov-Bohm
response through an area-dependent change in the surviving population.
Comparison with two-level dynamics identifies the interval in which
the adiabatic description remains accurate and shows its dependence on
the Raman coupling amplitude and imaginary-field
strength.

These
results provide signatures in the density, principal spreads, and
population of the surviving atoms. Spatial beam shaping and
state-dependent removal offer a route to observing these effects,
while interactions and controlled preparation of moving dressed
states remain directions for further study.

\section*{Acknowledgments}
This work was supported by the Research Council of Lithuania
(Grant No. S-ITP-24-6). The authors thank Ian B. Spielman, Tomoki Ozawa and Tomas Babelis for fruitful discussions. S.A. and D.B. used resources at the High Performance Computing Center (HPCC), ``HPC Saul\.etekis'' at Vilnius University, Faculty of Physics. D.B. also utilized the computing resources of the Scientific Computing and Data Analysis section of Core Facilities at Okinawa Institute of Science and Technology Graduate University (OIST).

\section*{Data Availability}
The numerical data underlying the figures are openly
available in Zenodo~\cite{AlminasData2026}. Code is available upon request from the corresponding authors.

\appendix

\section{Adiabatic projection of imaginary detuning Hamiltonian}
\label{app:adiabatic-proj}

\subsection{Two-level Hamiltonian and parametrization}
The two-level Hamiltonian of Eq.~\eqref{eq:H2-eff} is
\begin{equation}
 H_2=H_0\one+V_{\mathrm{at-l}}(\mathbf{r})
       -\frac{\ii\hbar\Gamma(\mathbf{r})}{2}\one,
 \label{eq:H2}
\end{equation}
where $H_0=\mathbf{p}^2/(2m)+V(\mathbf{r})$ and
$\mathbf{p}=-\ii\hbar\boldsymbol\nabla$. The coupling operator is
\begin{equation}
 V_{\mathrm{at-l}}=\frac{\hbar}{2}
 \begin{pmatrix}
 \ii\Delta&\Omega_R \ee^{-\ii\varphi}\\
 \Omega_R \ee^{\ii\varphi}&-\ii\Delta
 \end{pmatrix},
 \label{eq:Vat}
\end{equation}
with $\varphi=-2k_Ry$.

Here $m$ is the atomic mass, and $\Omega_R$, $\Delta$, and $\Gamma$ have units
of frequency. We take $\Omega_R>|\Delta|$ and real $\Delta$, so that
$V_{\mathrm{at-l}}$ has real, nondegenerate eigenenergies
\begin{equation}
 \varepsilon_s=\frac{s\hbar}{2}\sqrt{\Omega_R^2-\Delta^2},
 \qquad s=\pm1.
 \label{eq:energy}
\end{equation}
The common loss is kept separately from $\varepsilon_s$, as in the definition
of the dressed states in Eq.~\eqref{eq:dressed-eigen}.

Introduce the real hyperbolic parameter $\xi$ and a second real phase $\phi$:
\begin{equation}
 \begin{aligned}
 \tanh\xi&=\frac{\Delta}{\Omega_R},&
 \sinh\xi&=\frac{\Delta}{\sqrt{\Omega_R^2-\Delta^2}},\\
 \phi&=\arctan(\sinh\xi),&
 \boldsymbol\nabla\phi&=\frac{\boldsymbol\nabla\xi}{\cosh\xi}.
 \end{aligned}
 \label{eq:param}
\end{equation}
Thus $\varphi$ is the Raman phase, whereas $\phi$ is determined by $\xi$.

\subsection{Dressed states with unit right-right norm}
The right and left dressed states can be written in terms of the two phases as
\begin{align}
 \ket{\chi_s^{\mathrm R}}&=\frac{1}{\sqrt2}
 \left(\ee^{-\ii\varphi+\ii s\phi/2}\ket{\uparrow}
       +s \ee^{-\ii s\phi/2}\ket{\downarrow}\right),\label{eq:R}\\
 \ket{\chi_s^{\mathrm L}}&=\frac{\cosh\xi}{\sqrt2}
 \left(\ee^{-\ii\varphi-\ii s\phi/2}\ket{\uparrow}
       +s \ee^{\ii s\phi/2}\ket{\downarrow}\right).\label{eq:L}
\end{align}
They satisfy $V_{\mathrm{at-l}}\ket{\chi_s^{\mathrm R}}=\varepsilon_s\ket{\chi_s^{\mathrm R}}$ and
$V_{\mathrm{at-l}}^\dagger\ket{\chi_s^{\mathrm L}}=\varepsilon_s\ket{\chi_s^{\mathrm L}}$, with
\begin{equation}
 \begin{gathered}
 \braket{\chi_s^{\mathrm R}}{\chi_s^{\mathrm R}}=1,\qquad
 \braket{\chi_s^{\mathrm L}}{\chi_{s'}^{\mathrm R}}=\delta_{ss'},\\
 \braket{\chi_s^{\mathrm L}}{\chi_s^{\mathrm L}}=\cosh^2\xi.
 \end{gathered}
 \label{eq:norm}
\end{equation}
The two right eigenstates are generally not orthogonal:
$\braket{\chi_s^{\mathrm R}}{\chi_{-s}^{\mathrm R}}=-\ii s\tanh\xi$ in this phase convention.
For a single occupied dressed branch,
\begin{equation}
 \begin{gathered}
 \ket{\Psi_2(\mathbf{r},t)}=\psi_s(\mathbf{r},t)\ket{\chi_s^{\mathrm R}(\mathbf{r})},\\
 \braket{\Psi_2}{\Psi_2}=|\psi_s|^2.
 \end{gathered}
 \label{eq:density}
\end{equation}
The scalar coefficient therefore directly determines the physical density.

\subsection{Vector and scalar potentials}
Adiabatic projection with $\bra{\chi_s^{\mathrm L}}$ gives
$\ii\hbar\pd_t\psi_s=H_s\psi_s$, where
\begin{equation}
 H_s=\frac{(\mathbf{p}-\mathbf{A}_{s,s})^2}{2m}
       +V+W+\varepsilon_s-\frac{\ii\hbar\Gamma}{2},
 \label{eq:projected}
\end{equation}
where
\begin{equation}
 \begin{aligned}
 \mathbf{A}_{s,s'}&=\ii\hbar\braket{\chi_s^{\mathrm L}}{\boldsymbol\nabla\chi_{s'}^{\mathrm R}},\\
 W&=\frac{1}{2m}\mathbf{A}_{+,-}\mathbin{\cdot}\mathbf{A}_{-,+}.
 \end{aligned}
 \label{eq:potdef}
\end{equation}
The dot product contains no complex conjugation. 
The common loss from $H_2$
remains in Eq.~\eqref{eq:projected}; it is not included in $\varepsilon_s$.

Differentiating Eq.~\eqref{eq:R} and Eq.~\eqref{eq:L} yields
\begin{align}
 \braket{\chi_s^{\mathrm L}}{\boldsymbol\nabla\chi_s^{\mathrm R}}
 &=-\frac12\tanh\xi\,\boldsymbol\nabla\xi\notag\\
 &\quad-\frac{\ii}{2}(1+\ii s\sinh\xi)\boldsymbol\nabla\varphi,\label{eq:Ddiag}\\
 \braket{\chi_s^{\mathrm L}}{\boldsymbol\nabla\chi_{-s}^{\mathrm R}}
 &=-\frac{\ii\cosh\xi}{2}(\boldsymbol\nabla\varphi+s\boldsymbol\nabla\phi).
 \label{eq:Doff}
\end{align}
Consequently,
\begin{align}
 \mathbf{A}_{s,s}
 &=\frac{\hbar}{2}(1+\ii s\sinh\xi)\boldsymbol\nabla\varphi\notag\\
 &\quad-\frac{\ii\hbar}{2}\boldsymbol\nabla\ln(\cosh\xi),\label{eq:A}\\
 W&=\frac{\hbar^2\cosh^2\xi}{8m}
       \left[(\boldsymbol\nabla\varphi)^2-(\boldsymbol\nabla\phi)^2\right]\notag\\
  &=\frac{\hbar^2}{8m}
       \left[\cosh^2\xi\,(\boldsymbol\nabla\varphi)^2-(\boldsymbol\nabla\xi)^2\right].
 \label{eq:W}
\end{align}
The imaginary gradient in Eq.~\eqref{eq:A} follows from the position-dependent normalization of the right eigenstate. Rescaling a right state by $f(\mathbf r)$ and its left bra by $f^{-1}(\mathbf r)$ changes the connection by $\ii\hbar\nabla\ln f$; here $f=(\cosh\xi)^{-1/2}$ relative to a symmetric biorthogonal normalization. It has zero curl. The geometric scalar potential is real and
identical for the two branches. These expressions involve no expansion in
$\Delta/\Omega_R$; the approximation is the neglect of transitions between
dressed branches.

\subsection{Magnetic field}
On a region where the phases and dressed states are smooth,
$\boldsymbol\nabla\times\boldsymbol\nabla\varphi=0$ and
$\boldsymbol\nabla\times\boldsymbol\nabla\ln(\cosh\xi)=0$. Taking the curl of Eq.~\eqref{eq:A} gives
\begin{align}
 \mathbf{B}_s&=\boldsymbol\nabla\times\mathbf{A}_{s,s}\notag\\
 &=\frac{\ii s\hbar}{2}\boldsymbol\nabla(\sinh\xi)\times\boldsymbol\nabla\varphi\notag\\
 &=\frac{\ii s\hbar}{2}\cosh\xi\,\boldsymbol\nabla\xi\times\boldsymbol\nabla\varphi\notag\\
 &=\frac{\ii s\hbar}{2}\boldsymbol\nabla\!\left(
       \frac{\Delta}{\sqrt{\Omega_R^2-\Delta^2}}\right)
       \times\boldsymbol\nabla\varphi.
 \label{eq:B}
\end{align}
The field is purely imaginary, with opposite signs in the two branches. It
requires nonparallel gradients of the normalized detuning and the Raman phase.
Equivalently, directly in terms of the dressed states,
\begin{equation}
 (\mathbf{B}_s)_k=\ii\hbar\epsilon_{kij}
 \braket{\pd_i\chi_s^{\mathrm L}}{\pd_j\chi_s^{\mathrm R}}.
 \label{eq:Bstates}
\end{equation}
Here the second derivatives cancel under antisymmetrization; repeated Cartesian
indices are summed.

\subsection{Uniform-field configuration}
For $\varphi=-2k_Ry$ and $\xi=\xi(x)$, the vector potential and field are
\begin{align}
 \mathbf{A}_{s,s}&=A_x\mathbf{e}_x-\hbar k_R(1+\ii s\sinh\xi)\mathbf{e}_y,\notag\\
 A_x&=-\frac{\ii\hbar}{2}\tanh\xi\,\pd_x\xi,\label{eq:Aplanar}\\
 \mathbf{B}_s&=-\ii s\hbar k_R\frac{\dd}{\dd x}\left(
       \frac{\Delta}{\sqrt{\Omega_R^2-\Delta^2}}\right)\mathbf{e}_z.
 \label{eq:Bplanar}
\end{align}
A uniform field requires $\sinh\xi$ to be affine in $x$. Take
\begin{equation}
 \begin{aligned}
 \sinh\xi&=\eta x,& \Delta(x)&=\Omega_0\eta x,\\
 \Omega_R(x)&=\Omega_0\sqrt{1+\eta^2x^2},& \Gamma(x)&=\Gamma_0,
 \end{aligned}
 \label{eq:profiles}
\end{equation}
where $\Omega_0>0$ denotes the constant dressed-state splitting in frequency
units: $\varepsilon_s=s\hbar\Omega_0/2$. The hyperbolic parameter is
$\xi(x)=\operatorname{arsinh}(\eta x)$; $\eta$ has units of inverse length.

The resulting potentials and magnetic field are
\begin{align}
 A_x&=-\frac{\ii\hbar\eta^2x}{2(1+\eta^2x^2)},\label{eq:A_x-Apend}\\
 A_{y;s,s}&=-\hbar k_R-\ii s\Bcal x,\label{eq:Auniform}\\
 W(x)&=\frac{\hbar^2}{8m}\left[
    4k_R^2(1+\eta^2x^2)-\frac{\eta^2}{1+\eta^2x^2}\right],\notag\\
 \mathbf{B}_s&=-\ii s\Bcal\mathbf{e}_z,\qquad\Bcal=\hbar k_R\eta.
 \label{eq:Wuniform}
\end{align}
The constant real term $-\hbar k_R\mathbf{e}_y$ is removed by the phase choice
$\ket{\chi_s^{\mathrm{R,L}}}\mapsto \ee^{-\ii k_Ry}\ket{\chi_s^{\mathrm{R,L}}}$ and
$\psi_s\mapsto \ee^{\ii k_Ry}\psi_s$. Compensating the scalar potential with
$V=-W$ and omitting the constant eigenenergy then gives, for the lower branch,
\begin{equation}
 H=\frac{(p_x-A_x)^2+(p_y-\ii\Bcal x)^2}{2m}-\frac{\ii\hbar\Gamma_0}{2}.
 \label{eq:Hfinal}
\end{equation}
The denominator in $A_x$ is retained exactly; for $|\eta x|\ll1$,
$A_x\simeq-\ii\hbar\eta^2x/2$.

At constant Raman amplitude $\Omega_R=\Omega_0$, the same linear detuning
instead gives a nonuniform field,
\begin{equation}
 \mathbf{B}_s=-\ii s\frac{\hbar k_R\eta}{(1-\eta^2x^2)^{3/2}}\mathbf{e}_z,
 \qquad |\eta x|<1.
 \label{eq:constant}
\end{equation}
The formal divergence occurs at the closing gap, where the adiabatic
projection ceases to apply.

\begingroup
\color{black}
\section{Non-Hermitian Ehrenfest equation}
\label{app:ehr}

Consider an unnormalized state governed by
\begin{equation}
\ii\hbar\partial_t|\psi\rangle=H|\psi\rangle,\qquad
H=H_{\rm h}+\ii H_{\rm ah},
\label{eq:ehr-schrodinger}
\end{equation}
where $H_{\rm h}$ and $H_{\rm ah}$ are Hermitian. The normalized expectation
is $\langle O\rangle_t=\langle\psi|O|\psi\rangle/N(t)$, with
$N(t)=\langle\psi|\psi\rangle$. Differentiating the norm gives
\begin{equation}
\dot N=\frac{\ii}{\hbar}\langle\psi|(H^\dagger-H)|\psi\rangle
=\frac{2N}{\hbar}\langle H_{\rm ah}\rangle_t.
\label{eq:ehr-norm}
\end{equation}
For a time-independent observable, the numerator evolves as
\begin{equation}
\begin{aligned}
\frac{\dd}{\dd t}\langle\psi|O|\psi\rangle
&=\frac{\ii}{\hbar}\langle\psi|(H^\dagger O-OH)|\psi\rangle\\
&=\frac{\ii N}{\hbar}\langle[H_{\rm h},O]\rangle_t\\
&\quad+\frac{N}{\hbar}\langle\{H_{\rm ah},O\}\rangle_t.
\end{aligned}
\label{eq:ehr-numerator}
\end{equation}
The anti-Hermitian contributions from the bra and ket add, producing the
anticommutator. Differentiating the normalized expectation gives
\begin{equation}
\frac{\dd\langle O\rangle_t}{\dd t}
=\frac{1}{N}\frac{\dd}{\dd t}\langle\psi|O|\psi\rangle
-\frac{\dot N}{N}\langle O\rangle_t.
\label{eq:ehr-quotient}
\end{equation}
Substituting Eq.~\eqref{eq:ehr-norm} and Eq.~\eqref{eq:ehr-numerator}
yields Eq.~\eqref{eq:ehr} in the main text.
Its last term accounts for the changing norm.
A constant loss term $-\hbar\Gamma_0/2$ in $H_{\rm ah}$ cancels between
the last two terms and therefore does not affect normalized observables.
No Gaussian assumption is required in this derivation.
\endgroup

\section{Exact Gaussian dynamics and general initial conditions}
\label{app:gaussian}

\subsection{Transformations and inverted-oscillator propagator}

We solve the quadratic Hamiltonian in Eq.~\eqref{eq:Hcentral} for the
initial Gaussian in Eq.~\eqref{eq:psi0}, with
$\omega=\mathcal B/m=\hbar k_R\eta/m$.
The nonunitary substitution
\begin{equation}
\psi(\mathbf r,t)=\ee^{-\Gamma_0t/2}\ee^{\eta^2x^2/4}
\chi(\mathbf r,t)
\label{eq:prop-transform}
\end{equation}
gives $\ii\hbar\partial_t\chi=\widetilde H\chi$, where
\begin{equation}
\widetilde H=\frac{p_x^2+p_y^2}{2m}-\frac{m\omega^2x^2}{2}
-\ii\omega xp_y,
\label{eq:Htrans}
\end{equation}
and $\chi(\mathbf r,0)=\ee^{-\eta^2x^2/4}\psi_0(\mathbf r)$.
We use the Fourier convention
\begin{equation}
\chi(x,y,t)=\frac{1}{\sqrt{2\pi\hbar}}
\int_{\mathbb R}\dd p\,\ee^{\ii py/\hbar}\chi_p(x,t).
\label{eq:prop-Fourier}
\end{equation}
Since $[\widetilde H,p_y]=0$, each sector $p_y=p$ evolves independently.
Completing the square yields
\begin{equation}
\widetilde H_p=\frac{p_x^2}{2m}
-\frac{m\omega^2}{2}\left(x+\frac{\ii p}{m\omega}\right)^2.
\label{eq:prop-complete-square}
\end{equation}
The imaginary position shift is generated by the nonunitary operator
\begin{equation}
\mathcal T_p=\exp\left(-\frac{p\,p_x}{\hbar m\omega}\right),
\qquad
(\mathcal T_p f)(x)=f\!\left(x+\frac{\ii p}{m\omega}\right).
\label{eq:prop-shift-operator}
\end{equation}
Thus $\chi_p=\mathcal T_p\phi_p$, or equivalently
$\chi_p(x,t)=\phi_p(z,t)$ with $z=x+\ii p/(m\omega)$.
The transformed Hamiltonian is
\begin{equation}
\mathcal T_p^{-1}\widetilde H_p\mathcal T_p
=H_{\rm inv},\qquad
H_{\rm inv}=\frac{p_x^2}{2m}-\frac{m\omega^2x^2}{2}.
\label{eq:prop-inverted}
\end{equation}
Both transformations also act on the initial state:
\begin{equation}
\phi_p(z,0)=
\ee^{-\eta^2[z-\ii p/(m\omega)]^2/4}
\psi_{0,p}\!\left(z-\frac{\ii p}{m\omega}\right),
\label{eq:prop-initial}
\end{equation}
where $\psi_{0,p}$ is the Fourier transform of Eq.~\eqref{eq:psi0}.
The physical time-evolved envelope in each momentum sector is therefore
\begin{equation}
\begin{aligned}
\psi_p(x,t)&=\ee^{-\Gamma_0t/2}\ee^{\eta^2x^2/4}
\mathcal T_p\ee^{-\ii H_{\rm inv}t/\hbar}\\
&\quad\times\mathcal T_p^{-1}\ee^{-\eta^2x^2/4}\psi_{0,p}(x).
\end{aligned}
\label{eq:U}
\end{equation}
The physical density and moments are evaluated after restoring both
transformations.

The inverted-oscillator propagator is
\begin{equation}
\begin{aligned}
K_{\rm inv}(z,z';t)
&=\sqrt{\frac{m\omega}{2\pi\ii\hbar\sinh\omega t}}\\
&\quad\times\exp\!\left\{\frac{\ii m\omega}{2\hbar}
\left[\frac{z^2+z'^2}{\tanh\omega t}
-\frac{2zz'}{\sinh\omega t}\right]\right\}.
\end{aligned}
\label{eq:prop-inverted-kernel}
\end{equation}
The initial integration contour in the shifted coordinate is
$\mathbb R+\ii p/(m\omega)$. Setting
$z'=x'+\ii p/(m\omega)$ gives
\begin{equation}
\begin{aligned}
\psi_p(x,t)&=\ee^{-\Gamma_0t/2}\ee^{\eta^2x^2/4}
\int_{\mathbb R}\dd x'\,K_p(x,x';t)\\
&\quad\times\ee^{-\eta^2x'^2/4}\psi_{0,p}(x'),
\end{aligned}
\label{eq:prop-physical}
\end{equation}
\\
where
\begin{align}
K_p(x,x';t)
&=K_{\rm inv}\!\left(x+\frac{\ii p}{m\omega},
x'+\frac{\ii p}{m\omega};t\right)\notag\\
&=K_{\rm inv}(x,x';t)\notag\\
&\quad\times\exp\left\{\frac{\tanh(\omega t/2)}{\hbar}
\left[-p(x+x')-\frac{\ii p^2}{m\omega}\right]\right\}.
\label{eq:prop-shifted-kernel}
\end{align}
The physical envelope follows by the inverse Fourier transform of
$\psi_p$, using the convention of Eq.~\eqref{eq:prop-Fourier}.
For the initial state in Eq.~\eqref{eq:psi0}, both integrals are Gaussian.
Completing their squares and evaluating the normalized averages gives
the first moments, norm, and covariances below.

\subsection{General first moments and norm}
\label{app:gaussian-moments}

The position covariance is defined by
\begin{equation}
 v_{ij}(t)=\langle r_i r_j\rangle_t
 -\langle r_i\rangle_t\langle r_j\rangle_t,\qquad i,j\in\{x,y\}.
\end{equation}
At $t=0$, $v_{xx}(0)=\sigma_x^2$, $v_{yy}(0)=\sigma_y^2$,
and $v_{xy}(0)=0$.

To state the finite-time results compactly,
define
\begin{equation}
\begin{gathered}
\epsilon=\eta^2\sigma_x^2,\qquad
\delta=\frac{1+\epsilon}{4k_R^2\sigma_x^2},\\
c=\cosh\omega t,\qquad s=\sinh\omega t,
\end{gathered}
\label{eq:gauss-defs}
\end{equation}
and the denominators
\begin{align}
D_x&=1-[\epsilon+\delta(1+\epsilon)]s^2,\notag\\
\mathcal D_\eta&=D_x-\frac{1}{\sigma_y^2}
\left[\sigma_x^2s^2+(1+\epsilon)
\left(\frac{\hbar}{2m\omega\sigma_x}\right)^2(c-1)^2\right].
\label{eq:Dexact}
\end{align}

The first moments are given by
\begin{widetext}
\begin{subequations}\label{eq:mom}
\begin{align}
p_y(t)&=\frac{1}{\mathcal D_\eta}\left\{D_xp_{y0}
-\frac{\hbar}{2\sigma_y^2}\left([1-\delta(c-1)]x_0s
+\frac{p_{x0}}{m\omega}[c-1+\epsilon s^2]\right)\right\},\label{eq:Py}\\
x(t)&=\frac{1}{D_x}\left\{x_0c+(1+\epsilon)\frac{p_{x0}}{m\omega}s
-\frac{2}{\hbar}\left[\sigma_x^2c+(1+\epsilon)^2
\left(\frac{\hbar}{2m\omega\sigma_x}\right)^2(c-1)\right]sp_y(t)\right\},\label{eq:X}\\
p_x(t)&=\frac{1}{D_x}\left\{(1+\delta)m\omega x_0s+p_{x0}c
-\frac{2m\omega}{\hbar}\left[(1+\delta)\sigma_x^2s^2
+(1+\epsilon)\left(\frac{\hbar}{2m\omega\sigma_x}\right)^2c(c-1)\right]p_y(t)\right\},\label{eq:Px}\\
y(t)&=y_0+\frac{s}{m\omega}p_y(t)
-\frac{\hbar\eta^2}{2m\omega}(c-1)x(t).\label{eq:Ymom}
\end{align}
\end{subequations}
The surviving fraction is
\begin{equation}
N(t)=\frac{\ee^{-\Gamma_0t}}{\sqrt{\mathcal D_\eta}}
\exp\left\{\frac{\eta^2}{2(1+\epsilon)}
\left[\frac{[x_0c+(1+\epsilon)p_{x0}s/(m\omega)]^2}{D_x}-x_0^2\right]
+\frac{2\sigma_y^2}{\hbar^2}
\left[\frac{\mathcal D_\eta}{D_x}p_y^2(t)-p_{y0}^2\right]\right\}.
\label{eq:N}
\end{equation}
\end{widetext}

The terms proportional to $\sigma_x^2$ in Eq.~\eqref{eq:mom} describe
the effect of initial position uncertainty, whereas those proportional to
$1/\sigma_x^2$ describe initial momentum uncertainty. For the initial
Gaussian envelope, $\operatorname{Var}(p_x)_0=\hbar^2/(4\sigma_x^2)$,
so the factors $(\hbar/2m\omega\sigma_x)^2$ account for momentum
uncertainty.

The density remains Gaussian, with position covariance given by
Eq.~\eqref{eq:gaussian-covariance}, while the envelope momentum variance is
$\operatorname{Var}(p_y)=\hbar^2D_x/(4\sigma_y^2\mathcal D_\eta)$.
For a centered packet at rest, all four first moments remain zero,
although its covariance evolves and
$N(t)=\ee^{-\Gamma_0t}/\sqrt{\mathcal D_\eta}$.

We have checked initial-condition consistency: the general packet satisfies $N(0)=1$ and
\begin{equation}
\begin{aligned}
\dot x(0)&=\frac{(1+\epsilon)p_{x0}}{m}
-\frac{2\omega\sigma_x^2}{\hbar}p_{y0},&
\dot y(0)&=\frac{p_{y0}}{m},\\
\dot p_x(0)&=(1+\delta)m\omega^2x_0,&
\dot p_y(0)&=-\frac{\hbar\omega}{2\sigma_y^2}x_0.
\end{aligned}
\label{eq:slopes}
\end{equation}
The norm in Eq.~\eqref{eq:N} obeys Eq.~\eqref{eq:ehr-norm}.
Initially,
\begin{equation}
\left.\frac{\dd\ln N}{\dd t}\right|_{t=0}
=-\Gamma_0+\frac{\eta^2x_0p_{x0}}{m}
-\frac{2\omega x_0p_{y0}}{\hbar}.
\label{eq:norm-initial-slope}
\end{equation}

\subsection{Position covariance and principal directions}
\label{app:gaussian-covariances}

All position moments follow from the envelope density because
$\langle\chi_-^{\mathrm R}|\chi_-^{\mathrm R}\rangle=1$.
In particular, $x^2$, $xy$, and $y^2$ acquire no dressed-state
derivative corrections. The position covariance is
\begin{equation}
\mathbf v=\begin{pmatrix}
v_{xx}&v_{xy}\\v_{xy}&v_{yy}
\end{pmatrix},
\end{equation}
with entries defined in
Eq.~\eqref{eq:covdef}. Writing
$\mathbf q=\mathbf r-(x(t),y(t))^T$, the normalized density is
\begin{equation}
 \frac{|\psi(\mathbf r,t)|^2}{N(t)}=
 \frac{\exp[-\mathbf q^T\mathbf v^{-1}\mathbf q/2]}
 {2\pi\sqrt{\det\mathbf v}}.
 \label{eq:position-density}
\end{equation}
The same Gaussian integrals give the covariance below, with $\epsilon$, $\delta$, $c$, $s$, and
$\mathcal D_\eta$ as in Eq.~\eqref{eq:gauss-defs}
and Eq.~\eqref{eq:Dexact}.
\begin{subequations}\label{eq:gaussian-covariance}
\begin{align}
v_{xx}(t)&=\frac{1}{\mathcal D_\eta}
\biggl[\sigma_x^2c^2
+(1+\epsilon)^2\left(\frac{\hbar}{2m\omega\sigma_x}\right)^2s^2
\notag\\
&\hspace{10mm}-(1+\epsilon)
\left(\frac{\hbar}{2m\omega\sigma_y}\right)^2(c-1)^2\biggr],\\
v_{xy}(t)&=-\frac{\hbar(c-1)}{2m\omega\mathcal D_\eta}
\biggl[1+\epsilon+\frac{\sigma_x^2}{\sigma_y^2}(c+1)
\notag\\
&\hspace{8mm}+2(1+\epsilon)^2
\left(\frac{\hbar}{2m\omega\sigma_x\sigma_y}\right)^2(c-1)\biggr],\\
v_{yy}(t)&=\sigma_y^2+
\frac{1}{\mathcal D_\eta}
\left(\frac{\hbar}{2m\omega\sigma_y}\right)^2
\notag\\
&\hspace{8mm}\times\left[s^2+(1+\epsilon)
(\eta^2\sigma_y^2-4\delta)(c-1)^2\right].
\end{align}
\end{subequations}
The result is valid on the normalizable branch
$\mathcal D_\eta>0$ connected to $t=0$. It depends on the initial
widths but not on the initial position or momentum. Uniform loss
cancels from the normalized covariance. The laboratory standard deviations are
$\sigma_x(t)=\Delta x=\sqrt{v_{xx}(t)}$ and
$\sigma_y(t)=\Delta y=\sqrt{v_{yy}(t)}$,
and the uncentered second moments are
$\langle r_i r_j\rangle_t=v_{ij}(t)+r_i(t)r_j(t)$.

The eigenvalues of $\mathbf v$ are
\begin{equation}
 \lambda_\pm=\frac{v_{xx}+v_{yy}}{2}
 \pm\frac12\sqrt{(v_{xx}-v_{yy})^2+4v_{xy}^2}.
 \label{eq:principal-variances}
\end{equation}
For distinct eigenvalues, the corresponding unit vectors are
\begin{equation}
 \begin{aligned}
 \theta_+&=\tfrac12\operatorname{atan2}
 (2v_{xy},v_{xx}-v_{yy}),\\
 \mathbf e_+&=(\cos\theta_+,\sin\theta_+)^T,\qquad
 \mathbf e_-=(-\sin\theta_+,\cos\theta_+)^T.
 \end{aligned}
 \label{eq:principal-directions}
\end{equation}
The two-argument arctangent fixes the quadrant. Since the variance
along any unit vector $\mathbf n$ is
$\mathbf n^T\mathbf v\mathbf n$, the maximum and minimum
widths are $\sqrt{\lambda_+}$ along $\mathbf e_+$ and
$\sqrt{\lambda_-}$ along $\mathbf e_-$. These are also the axes of
the Gaussian density contours. The eigenvalues are variances, rather
than widths. When they coincide, no direction is preferred.

The initial covariance is
$\mathbf v(0)=\operatorname{diag}(\sigma_x^2,\sigma_y^2)$.
For the unchirped initial state, its change begins at order $t^2$,
with
\begin{equation}
 v_{xy}(t)=-\frac{\hbar\omega}{4m}
 \left(1+\epsilon+2\frac{\sigma_x^2}{\sigma_y^2}\right)t^2+O(t^4).
 \label{eq:covariance-short-time}
\end{equation}
Reversing the field reverses this shear and reflects the principal
axes, while leaving their widths unchanged. As $\eta\to0$ at fixed
$k_R$, both vector-potential components vanish and the covariance
reduces to free expansion,
$v_{ii}=\sigma_i^2+\hbar^2t^2/(4m^2\sigma_i^2)$ and
$v_{xy}=0$.

\subsection{Principal spreads for the circular launch}
\label{app:circular-launch}

For Eq.~\eqref{eq:y-launch}, set $\sigma_x=\sigma_y=\sigma$ in
Eq.~\eqref{eq:mom}, Eq.~\eqref{eq:N}, and Eq.~\eqref{eq:gaussian-covariance}.
The short-time first moments give Eq.~\eqref{eq:y-launch-drift_x}, Eq.~\eqref{eq:y-launch-drift_y}, and
$p_y(t)=p_{y0}[1+\omega^2t^2]+O(t^4)$.
To obtain the principal-width criterion, write
$r=\hbar/(2m\omega\sigma^2)$ and $\epsilon=\eta^2\sigma^2$.
The covariance expands as
\begin{equation}
 \begin{aligned}
 \frac{\mathbf v(t)}{\sigma^2}
 ={}&\mathbf1_2+(\omega t)^2
 \begin{pmatrix}
 2+\epsilon+(1+\epsilon)^3r^2 & -(3+\epsilon)r/2\\
 -(3+\epsilon)r/2 & r^2
 \end{pmatrix}\\
 &+O(t^4).
 \end{aligned}
 \label{eq:circular-covariance-expansion}
\end{equation}
The matrix multiplying $(\omega t)^2$ has positive trace and determinant
$r^2(1+\epsilon)^2[(1+\epsilon)r^2-1/4]$.
Its smaller eigenvalue is negative precisely when
$4(1+\epsilon)r^2<1$, which gives Eq.~\eqref{eq:y-launch-narrowing}.
At equality, substitution into the finite-time covariance gives
\begin{equation}
 \begin{aligned}
 \lambda_-(t)&=\sigma^2,\\
 \lambda_+(t)&=\sigma^2\left[1+
 \left(\frac{(3+\epsilon)^2}{4}+r^2\right)
 \frac{s^2}{\mathcal D_\eta}\right],\\
 \theta_+(t)&=-\arctan\frac{2r}{3+\epsilon}.
 \end{aligned}
 \label{eq:circular-fixed-axes}
\end{equation}
Here $\mathcal D_\eta$ is evaluated at equal initial widths.

At $t=0$, the angle denotes its limiting value because the initial
covariance is degenerate. For other parameter values, the exact
eigenvalues and directions follow from
Eq.~\eqref{eq:principal-variances} and Eq.~\eqref{eq:principal-directions}.

\bibliographystyle{apsrev4-2}
\bibliography{references}

\end{document}